\documentclass[preprint,12pt]{elsarticle}

\usepackage[
left=1.0in,
right=1.0in,
top=0.85in,
bottom=0.9in
]{geometry}

\usepackage{eulervm}
\usepackage{amsmath}
\usepackage{amssymb}
\usepackage{bm}
\usepackage{dutchcal}
\usepackage{mathrsfs}
\usepackage{caption,subfig} % subfloat in figure
\usepackage{float}
\usepackage{array}
\usepackage{amssymb}
\usepackage{mathtools}
\usepackage{amsmath,amsfonts,bm}
\usepackage{enumerate}
\usepackage[table]{xcolor}
\usepackage{enumerate}
\usepackage{enumitem}
\usepackage{microtype}
\usepackage{booktabs}
\usepackage{array}
\usepackage{float}
\usepackage{url}
\usepackage[hidelinks]{hyperref}
\usepackage{setspace}
	\usepackage{empheq}
\setlist{nosep,leftmargin=*}

\newtheorem{defn}{Definition}

\newtheorem{thm}{Theorem}
\newtheorem{lemma}{Lemma}
\newtheorem{corollary}{Corollary}

\newtheorem{remark}{Remark}
 \newtheorem{fact}{Fact}
\newtheorem{exm}{Example}
\makeatletter
\@ifundefined{proof}{}{%

}
\makeatother

\newenvironment{proof}
{\par\noindent\textbf{Proof.}\ }
{\hfill$\square$\par}

\newcommand{\PoA}{\mathrm{PoA}}

\DeclareMathOperator*{\argmin}{argmin}

\begin{document}
	
	\begin{frontmatter}
		
		\title{On the Tragedy of the Commons in Heterogeneous Distributed Resource Allocation }
		
		\author[label1]{Avadh Kishor\corref{cor1}}
		\ead{akishor@iiitm.ac.in} 
		
		\author[label2]{Rajdeep Niyogi}
		\ead{rajdeep.niyogi@cs.iitr.ac.in}
		
		\address[label1]{Department of Computer Science and Engineering\\ABV-Indian Institute of Information Technology and Management Gwalior, India 474015}
			\address[label2]{Department of Computer Science and Engineering\\Indian Institute of Technology Roorkee, Roorkee, India 247667}
		
		\cortext[cor1]{Corresponding author}

		\begin{abstract}
			
				We consider the Tragedy of the Commons (ToC) problem, which commonly arises in heterogeneous distributed resource allocation, in which selfish users are independently interested in optimizing their own utility. Such self-interested decision-making in decentralized settings often results in inefficient resource utilization because users prefer high-capacity resources, leading to their overutilization. We cast the problem into a non-cooperative game and derive closed-form characterizations of both the distributed Nash equilibrium (NE) and the socially optimal allocation. In view of these closed-form expressions, we establish the structural conditions under which the ToC arises, and derive an exact
			expression for the Price of Anarchy (PoA). We further prove that the PoA
			can become arbitrarily large (unbounded) under suitable traffic conditions. Motivated by these observations, we propose a
			penalty-based incentive mechanism that discourages the use of
			overloaded servers while preserving decentralized decision-making. 
We analyze the resulting equilibrium load
redistribution and establish conditions under
which the proposed mechanism reduces the overall
system cost and improves equilibrium efficiency.

		\end{abstract}
		
		\begin{keyword}
		Resource allocation\sep Non-cooperative game\sep Nash equilibrium\sep Tragedy of the commons\sep Price of anarchy\sep Distributed systems 
		\end{keyword}
		
	\end{frontmatter}

\section{Introduction}

Resource allocation is a fundamental problem in distributed computing,
cloud systems, communication networks, and large-scale multi-agent systems, where a collection of autonomous users competes for limited
shared resources. Managing resource allocation efficiently directly affects
important performance measures such as response time, throughput,
resource utilization, energy consumption, and quality of service. The classical way of dealing with these problems is to use centralized
optimization techniques that assume complete knowledge of the system and a central controller that can compute global optimal allocations.
Although such approaches often achieve high system efficiency, they
become increasingly difficult to deploy in modern distributed
environments, where resources are geographically distributed, users act
independently, and centralized coordination is either computationally
expensive or practically infeasible. Consequently, decentralized resource allocation has emerged as an important research direction in which each user independently selects resources according to their own objective without requiring global system information \cite{nisan2007algorithmic,harks2022equilibrium}.

Game theory provides a mathematical framework for analyzing such decentralized systems. In particular, non-cooperative game models capture selfish behavior and the inherent competition to optimize utility. 
A strategy profile in which no player can benefit from a unilateral deviation is called the Nash equilibrium (NE). With this view, game-theoretic solutions for resource allocation have therefore been extensively applied to distributed computing environments \cite{shamshirband2021survey,grosu2005noncooperative,penmatsa2011static}. However, selfish decision-making generally
optimizes individual performance rather than the overall system
objective.

One of the most fundamental manifestations of the conflict between the individual and the system as a whole is the Tragedy of the Commons (ToC), which describes a situation in which myopic, selfish, rational users overlook the well-being of the system as a whole, resulting in the system's doom. Interestingly, a similar phenomenon can also be observed in distributed computing and communication systems, where all rational users prefer faster and less expensive resources to optimize their quality of service. Consequently, some resources (high-end) become overloaded, while others (slower servers) remain underutilized \cite{lopez2005mathematical,saha2003local,kishor2016altruistic}. Ultimately, this equilibrium phenomenon yields inefficient results compared to the centrally coordinated optimum. 

The efficiency loss due to selfish behavior is quantified as the worst-case NE and that of the socially optimal allocation, and is known as the Price of Anarchy (PoA)~~\cite{roughgarden2002bad,christodoulou2005price}. PoA has been extensively investigated in distributed resource allocation in various domains
\cite{nisan2007algorithmic,chen2020randomscheduler,paccagnan2019utility}. These results provide valuable
performance guarantees and establish the theoretical limits of selfish
systems. Nevertheless, they mainly quantify the magnitude of the
inefficiency and provide comparatively limited insight into the
structural mechanisms responsible for its occurrence. In particular,
they do not explicitly explain why selfish users overload specific
resources, how the resulting equilibrium differs structurally from the
socially optimal allocation, or how these structural differences can be
analytically exploited to design effective mitigation mechanisms.

These observations motivate a different perspective on the analysis of
selfish resource allocation. Instead of viewing the ToC solely as a consequence of selfish decision making, we regard it
as a structural property of heterogeneous distributed resource
allocation. Our objective is therefore not only to quantify the
inefficiency through the PoA but also to identify the
fundamental structural conditions that give rise to this phenomenon.
Specifically, we seek to answer several natural theoretical questions.
Under what conditions does selfish routing overload high-capacity
servers? How does the distributed equilibrium differ from the socially
optimal allocation? Can the inefficiency be characterized through exact closed-form expressions rather than worst-case bounds? More importantly, can an incentive mechanism be designed to reduce the inefficiency of the distributed equilibrium?

The above questions are particularly important for heterogeneous
distributed systems, including cloud data centers, edge computing
platforms, communication networks, and shared computing infrastructures, where servers often possess different service capacities, and users make completely decentralized decisions. Although extensive research has been devoted to equilibrium computation and PoA analysis, there is a lack of a unified analytical framework that simultaneously characterizes the distributed equilibrium, the socially optimal allocation, the structural origin of the ToC, and its exact efficiency loss. Thus, addressing this gap constitutes the primary motivation of the present work.

\subsection{Our Contribution}

We adopt the view that understanding the structural
origin of the ToC is essential for developing
effective mitigation mechanisms. Instead of treating
the PoA merely as a worst-case performance measure,
we investigate the relationship between the
distributed Nash equilibrium and the socially
optimal allocation to identify the structural
conditions that lead to resource overloading.
This perspective enables us to derive exact
analytical characterizations of equilibrium
inefficiency and to establish how this inefficiency
varies with the system parameters. Building upon
these theoretical results, we propose a penalty-based
incentive mechanism and analyze its effect on
equilibrium load distribution, overall system cost,
and the PoA.

The main contributions of this paper are summarized
as follows.

\begin{itemize}
	
	\item[$\triangleright$]
	We model selfish allocation among heterogeneous
	$M/M/1$ servers as an atomic splittable
	non-cooperative game. We obtain each user's unique
	best response, derive the coupled Nash equilibrium
	conditions, and characterize a symmetric-user
	equilibrium. We also obtain the centralized optimum
	in terms of server capacities and total incoming
	load (Section~\ref{sec:characterization}).
	
	\item[$\triangleright$]
	We investigate the structural origin of the ToC
	in heterogeneous distributed systems. In particular,
	Theorem~\ref{thm:main_inefficiency} establishes
	that the number of servers utilized in a distributed
	equilibrium is no greater than that of the socially
	optimal allocation, and that the distributed
	equilibrium assigns more load to high-capacity
	servers.
	
	\item[$\triangleright$]
	We derive an exact expression for the equilibrium
	cost ratio and show that it depends on the
	equilibrium and socially optimal server loads
	rather than on the complete routing profile
	(Theorem~\ref{thm:exactpoa}).
	
	\item[$\triangleright$]
	We analyze the asymptotic behavior of the PoA
	and establish that it is unbounded for a suitable
	family of instances
	(Theorem~\ref{thm:unboundedpoa}).
	
	\item[$\triangleright$]
	We propose a penalty-based incentive mechanism
	that discourages users from selecting overloaded
	servers. For symmetric users, we establish
	conditions under which the mechanism redistributes
	load from high-capacity servers to lower-capacity
	servers (Theorem~\ref{thm:final}).
	
	\item[$\triangleright$]
	Finally, we establish sufficient conditions under
	which the proposed penalty mechanism decreases
	the overall system cost
	(Lemma~\ref{lem:pen-cost-reduction}) and improves
	the equilibrium cost ratio
	(Theorem~\ref{thm:penalty-poa}).
	
\end{itemize}

The remainder of this paper is organized as follows. The review of the existing literature is discussed in Section~\ref{sec:rel}. In Section~\ref{sec:prelim}, the preliminaries, including the heterogeneous distributed resource allocation model and the formalization of the distributed and socially optimal resource allocation, are presented.
Section~\ref{sec:characterization} derives the distributed Nash equilibrium
and the socially optimal allocation, together with their analytical
characterizations. Then, the structural origin of ToC is established in Section~\ref{sec:esttoc}. The analytical characterization of PoA is given in  
Section~\ref{sec:quantpoa}. The penalty-based incentive mechanism Section~\ref{sec:soltoc} followed by theoretical results on equilibrium load redistribution~\ref{sec:penaltypoa}. Finally,
Section~\ref{sec:con} concludes the paper and discusses possible
directions for future research.

\section{Related Work}\label{sec:rel}

Distributed resource allocation is the problem of finding allocation
strategies for autonomous users competing for shared resources in a
decentralized environment. Traditionally, the simplest way to deal with such problems is centralized scheduling. Although centralized approaches often yield globally efficient solutions, the requirement of complete system information makes them impractical for large-scale distributed environments. Consequently, increasing attention has been devoted to decentralized resource allocation, where resource sharing is modeled as a strategic interaction among selfish agents
\cite{jiang2015survey,harks2022equilibrium}. Game-theoretic models are widely used for decentralized allocation~\cite{shamshirband2021survey}. Beyond computing systems, game-theoretic models have also been applied to resource-sharing problems in economics and social systems. For example, Gamez
\emph{et al.}~\cite{gamez2018marketing} developed an evolutionary
game-theoretic framework for cooperative fisheries to analyze strategic interactions among participants. Although these studies demonstrate the versatility of game theory in modeling decentralized decision-making, their primary emphasis is on equilibrium computation or application-specific optimization rather than understanding the inefficiency introduced by selfish resource allocation.

A closely related line of research concerns non-cooperative load
balancing in heterogeneous distributed systems. Grosu and
Chronopoulos~\cite{grosu2005noncooperative} formulated distributed load balancing as a non-cooperative game, and analyzed the existence of stable allocations under selfish behavior. They subsequently introduced a
truthful mechanism that encourages users to reveal their actual
requirements while improving fairness in a distributed task
allocation~\cite{grosu2004mechanism}. Penmatsa and
Chronopoulos~\cite{penmatsa2011static} further investigated both static
and dynamic game-theoretic load balancing algorithms for heterogeneous
distributed systems and demonstrated that the Nash equilibrium-based resource allocation achieves better scalability than centralized
approaches. 
More recently, Polgar and
Varga~\cite{polgar2023loadbalancing} proposed game-theoretic load
balancing algorithms for heterogeneous wireless backhaul networks,
showing that auction-based mechanisms substantially outperform purely
selfish routing strategies. Similarly, several distributed resources
allocation schemes have been proposed for cloud and heterogeneous
computing environments
\cite{avadh20,KISHOR202029}. While these studies provide efficient
distributed allocation algorithms under selfish users, they primarily
focus on equilibrium computation and algorithmic performance rather than
providing a theoretical characterization of the Tragedy of the Commons
or its impact on equilibrium efficiency.

Another important research direction studies the inefficiency introduced
by decentralized decision-making. For example, Chen
\emph{et al.}~\cite{chen2020randomscheduler} analyzed load balancing
games under randomizing schedulers and defined bounds for both PoA and the Price of Stability under several behavioral models. Their results demonstrated that equilibrium efficiency depends strongly on users' decision-making principles and risk preferences. Paccagnan \emph{et al.}~\cite{paccagnan2019utility} investigated utility design for distributed resource allocation and proposed utility functions that optimize the exact PoA, thereby providing
performance guarantees for decentralized systems. Although these studies offer valuable insights into equilibrium efficiency, their primary objective is to quantify or optimize the worst-case performance of selfish equilibria rather than to identify the structural causes of inefficiency or the emergence of congestion under heterogeneous resources.

 Early studies investigated the emergence of the ToC in
distributed artificial intelligence and multi-agent resource sharing,
showing that selfish agents may collectively overexploit common
resources, even though such behavior is detrimental to the overall
system~\cite{turner1993tragedy}. To alleviate this phenomenon, researchers have proposed a variety of decentralized coordination
mechanisms. Mundhe and Sen~\cite{mundhe2000evolving} employed genetic algorithms to evolve cooperative multi-agent societies, while
Tumer and Wolpert~\cite{tumer2000collective} introduced the Collective
Intelligence (COIN) framework to align local objectives with global
performance. Saha \emph{et al.}~\cite{saha2003local} proposed a
local-feedback mechanism for mitigating the ToC in multi-agent systems,
whereas Kishor \emph{et al.}~\cite{kishor2016altruistic} developed an altruistic decision-making algorithm that dynamically adjusts agents'
motivation and eagerness to reduce selfish behavior. Similar phenomena
have also been identified in communication networks, where Lopez
\emph{et al.}~\cite{lopez2005mathematical} established a mathematical framework demonstrating that TCP congestion control may exhibit ToC and proposed protocol modifications to alleviate its adverse effects. These studies convincingly demonstrate the occurrence of the ToC in distributed systems, but mainly focus on algorithmic remedies rather than deriving analytical characterizations of the underlying inefficiency.

Further,  Platkowski~\cite{platkowski2016egalitarian} proposed an egalitarian payoff redistribution mechanism for multiplayer social dilemma games and showed that equitable redistribution substantially promotes long-term cooperation. Subsequently, Platkowski
\cite{platkowski2017greed} introduced a generalized mathematical model
capturing greed and fear in multiplayer social dilemmas and
demonstrated that these two factors fundamentally govern the evolution
of selfish behavior in public goods games and ToC scenarios. Numerous cooperation-enhancing mechanisms have since been
proposed. He \emph{et al.}~\cite{he2019monitoring} showed that
institutional monitoring together with individual reporting
significantly suppresses free-riding in collective-risk social
dilemmas. Gao \emph{et al.}~\cite{gao2019redistribution} demonstrated
that asymmetric payoff redistribution effectively promotes cooperation. In contrast, Jiang \emph{et al.}~\cite{jiang2021bystander} established that reducing the bystander effect through smaller interaction groups leads
to higher cooperation levels. More recently, Su \emph{et al.}~\cite{su2023interdependent} investigated cooperation on
interdependent networks and showed that coupling between network layers enhances network reciprocity and significantly improves cooperative behavior. Although these works provide valuable insights into the evolution of cooperation based on simulation analyses, they did not consider the characterization of equilibrium efficiency.

Another important research direction introduces incentive mechanisms to
improve resource utilization while preserving decentralized
decision-making. Mehta and Vazirani~\cite{mehta2020market} proposed a market-based mechanism for air traffic flow management that achieves
efficient and incentive-compatible allocation through the market
equilibrium. Chen \emph{et al.}~\cite{chen2021fair} developed an
envy-free max--min fairness mechanism based on coalitional game theory, providing both efficiency and fairness guarantees for resource allocation
allocation. More recently, Gao \emph{et al.}~\cite{gao2025ecological}
introduced an ecological compensation mechanism for the common pool
resource games, demonstrating that appropriate compensation intensity
promotes cooperation, sustainable resource utilization, and higher
population welfare. Similarly, Lee
\emph{et al.}~\cite{lee2023myopic} showed that introducing myopic
decision-making, together with additional intermediate strategies
substantially restores cooperation in structured social dilemma games.
These studies demonstrate that pricing, compensation, and incentive
mechanisms are effective tools for encouraging cooperation. However,
their primary objective is to improve evolutionary outcomes rather than
provide analytical guarantees on equilibrium efficiency or the PoA.

A related approach uses tolls and coordination
mechanisms to influence selfish routing decisions.
Christodoulou et al.~\cite{christodoulou2014improving}
studied coordination mechanisms for selfish routing
and their effect on the PoA. However, their results
cannot be directly applied to our model, where
users distribute their loads among heterogeneous
$M/M/1$ servers. Unlike nonatomic routing, each
user's allocation affects server congestion.
Therefore, our analysis considers the individual
optimality conditions of an atomic splittable
resource allocation game.

Despite these significant advances, several important theoretical
questions remain open. Existing studies primarily focus on equilibrium
computation, cooperation promotion through evolutionary dynamics,
payoff redistribution, monitoring, punishment, ecological compensation,
or pricing mechanisms. Comparatively little attention has been devoted
to developing a unified analytical framework for heterogeneous
distributed resource allocation under selfish users. In particular, the
existing literature lacks theoretical results that simultaneously
characterize the socially optimal allocation, distributed Nash
equilibrium, structural origin of the ToC,  exact
closed-form expression of the PoA, its asymptotic behavior,
and provably effective incentive mechanisms within a single analytical
framework. The present work addresses these challenges by deriving
closed-form characterizations of both the distributed equilibrium and
the socially optimal allocation, establishing the structural conditions
under which the ToC arises, obtaining an exact
expression for the PoA together with its asymptotic
behavior, and finally proposing a penalty-based incentive mechanism
that provably reduces congestion on overloaded servers, decreases the overall system cost, improves the PoA, and consequently
mitigates the ToC while preserving decentralized decision making.

\section{Preliminaries} \label{sec:prelim}

\subsection{Distributed Resource Allocation Model} \label{subsec:distmodel}
\begin{figure}[htb!]
	\centering
	\includegraphics[width= 13cm, height = 6cm] {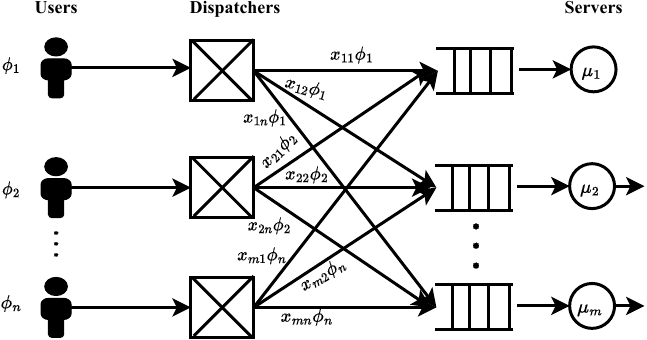} 
	
	\caption{An architecture demonstrating a distributed resource allocation model}
	\label{fig:decentral}
	
\end{figure}

 We consider a distributed computing environment consisting of a set of users, each located at a distinct geographical location, and a set of heterogeneous servers capable of processing incoming jobs. Following prior work~\cite {brun2016worst,KISHOR202029}, we consider a prevalent distributed system architecture comprising three components: users, dispatchers, and servers, as illustrated in Figure~\ref{fig:decentral}. Let $[n]=\{1,2,\dots,n\}$ be a set of users. Each user $i\in[n]$ submits its jobs to a dedicated dispatcher, resulting in a one-to-one mapping between users and dispatchers. Users are co-located with their respective dispatchers, and the underlying load allocation mechanism is abstracted away from the users. Each dispatcher acts on behalf of its associated user by routing incoming jobs to a shared set of servers $[m]=\{1,2,\dots,m\}$. Each server $j\in[m]$ is characterized by its processing rate $\mu_j$. The result of each processed job is returned from the server to the corresponding dispatcher, who then forwards it to the user.

%%%%%%%%%%%%%%%%%%%%%%%%%%%%%%%%%%%

Since dispatchers act as logical proxies for users and make no independent strategic choices, we henceforth use the terms \emph{user} and \emph{dispatcher} interchangeably. Likewise, we interchange \emph{job} and \emph{load} throughout the paper. Each user $i\in[n]$ generates load at a rate $\phi_i$ (measured in jobs per unit time) and determines how to distribute this load across the servers. The load allocation strategy of user $i$ is specified by a vector $\mathbf{x}_i=(x_{i1},x_{i2},\dots,x_{im})\in[0,1]^m$, where $x_{ij}$ denotes the fraction of $\phi_i$ routed to server $j$. Further, the strategy $\mathbf{x}_i$ is feasible if and only if the following conditions hold: 
\begin{enumerate}[label=(\roman*)]
	\item a user assigns a non-negative amount of load to a server, i.e., $x_{ij}\geq 0$,
	\item a user assigns its jobs to at least one server, i.e., $\sum_{j=1}^{m}x_{ij}=1$, $\forall i\in[n]$,
	\item the total job arrival rate on a server must be less than its processing rate, i.e., $\sum_{i=1}^{n}x_{ij}\phi_i<\mu_j$, $\forall j\in[m]$.
\end{enumerate}

 We assume that the jobs generated by user $i$ form a Poisson process of rate $\phi_i$. Each job is independently routed to exactly one server $j$ with probability $x_{ij}$. Therefore, by Poisson thinning and superposition, server $j$ receives a Poisson stream of aggregate rate
$
\lambda_j=\sum_{i=1}^n\phi_i x_{ij}.
$
Server $j$ has independent exponential service times of rate $\mu_j$ and operates as an $M/M/1$-FCFS queue.

According to Little's Law, the average response time in a queuing system is directly related to the average number of tasks in the system and their arrival rate. As established in~\cite{DBLP:books/daglib/0079988}, consider a feasible strategy profile $\mathbf{X}=(\mathbf{x}_i,\mathbf{x}_{-i})$, where $\mathbf{x}_{-i}$ denotes the strategies of all dispatchers except dispatcher $i$. The expected response time at server $j\in[m]$ is given by
\begin{equation}\label{eq:srt}
	\mathcal{D}_j(\mathbf{x}_i,\mathbf{x}_{-i})
	=\frac{1}{\mu_j-\sum_{i=1}^{n}x_{ij}\phi_i}.
\end{equation}

The contribution of server $j$ to the mean response time of a randomly selected job of user $i$, rather than the completion time of parallel subtasks of the same job, is
\begin{equation}\label{eq:drt1}
	\mathcal{R}_i^j(\mathbf{x}_i,\mathbf{x}_{-i})
	=x_{ij}\mathcal{D}_j(\mathbf{x}_i,\mathbf{x}_{-i})
	=\frac{x_{ij}}{\mu_j-\sum_{k=1}^n x_{kj}\phi_k}.
\end{equation}

By conditioning on the server to which a job is routed, the law of total expectation gives the expected response time for dispatcher $i$ as
\begin{equation}\label{eq:drt}
	\mathcal{R}_i(\mathbf{x}_i,\mathbf{x}_{-i})
	=\sum_{j=1}^{m}\mathcal{R}_i^j(\mathbf{x}_i,\mathbf{x}_{-i})
	 =\sum_{j=1}^{m}\frac{x_{ij}}{\mu_j^i-x_{ij}\phi_i},
\end{equation}
where the residual capacity available to dispatcher $i$ at server $j$ is defined as
\begin{equation}\label{eq:availcap}
	\mu_j^i
	=\mu_j-\sum_{\substack{k=1\\k\neq i}}^n x_{kj}\phi_k.
\end{equation}

\paragraph{Optimization Problem:}We consider dispatcher $i$'s distributed  (individual) objective, which seeks to minimize its expected response time, assuming $\mathbf{x}_{-i}$ is fixed. The corresponding optimization problem of dispatcher $i$ (labeled as $\mathsf{DISOPTIM}_i)$ is:
\begin{align}
	(\mathsf{DISOPTIM}_i) \qquad \min_{\mathbf{x}_i \in [0,1]^m} \quad & \mathcal{R}_i(\mathbf{x}_i, \mathbf{x}_{-i}) \label{eq:pb} \\
	\text{s.t.} \quad & \sum_{j=1}^{m} x_{ij} = 1 \label{eq:con1} \\
	& x_{ij} \geq 0 \qquad \forall j \in [m] \label{eq:con2} \\
	& x_{ij} \phi_i < \mu_j^i \qquad \forall j \in [m], \label{eq:con3}
\end{align}

\subsection{Centralized Resource Allocation Model}

The centralized model provides a benchmark for measuring the loss caused by selfish decisions. We consider the same system setting as in the distributed model, with $[n]$ users and $[m]$ heterogeneous servers. However, in contrast to the distributed model, a single centralized dispatcher routes all user loads to the servers. Each user $i$ generates load at rate $\phi_i>0$, and each server $j$ has service capacity $\mu_j>0$. Let $\Phi=\sum_{i=1}^{n}\phi_i$ denote the total incoming load to the system. The centralized dispatcher chooses a load allocation strategy $\mathbf{s}=(s_1,\dots,s_m)$, where $s_j\in[0,1]$ denotes the fraction of load routed to server $j$. A load assignment is \emph{feasible} if (i) $\sum_{j=1}^{m}s_j=1$, and (ii) $s_j\Phi<\mu_j$ for every $j\in[m]$. The corresponding aggregate load at server $j$ is $\lambda_j^s=s_j\Phi$.

Considering that each server operates as an $M/M/1$-FCFS queue, the system-wide expected response time (social cost) is given by
\begin{equation}\label{eq:crt}
	\mathcal{C}(\mathbf{s})
	=\sum_{j=1}^{m}\frac{s_j}{\mu_j-s_j\Phi}.
\end{equation}
%For comparison with Section~6, the total delay workload is $C(\pi_s)=\Phi\mathcal{C}(\mathbf{s})$.

\paragraph{Optimization Problem:}
The socially optimal allocation minimizes the system-wide expected response time subject to feasibility constraints. The corresponding optimization problem, labeled as $\mathsf{SOCOPTIM}$, is
\begin{align}
	(\mathsf{SOCOPTIM}) \qquad
	\min_{{s_j}\in[0,1]}\mathcal{C}(\mathbf{s})
	&=\sum_{j=1}^{m}\frac{s_j}{\mu_j-s_j\Phi},
	\label{eq:cpb}\\
	\text{s.t.}\quad \sum_{j=1}^{m}s_j
	&=1, \label{eq:ccon1}\\
	s_j&\geq0,\qquad \forall j\in[m],
	\label{eq:ccon2}\\
	s_j\Phi&<\mu_j,\qquad \forall j\in[m].
	\label{eq:ccon3}
\end{align}

\section{Characterization of Optimal Strategies in Distributed and Centralized Models}\label{sec:characterization}
In this section, we compute the optimal allocation strategies for 
distributed and centralized resource allocation schemes. Specifically, we first formulate the distributed resource allocation setting as a non-cooperative game and characterize its Nash equilibrium. We then compute the optimal allocation strategies for the centralized setting by solving a global optimization problem. 

\subsection{Distributed Allocation as a Non-cooperative Game}

In the distributed setting, each user (dispatcher) independently selects how to distribute its incoming job flow across the available servers. The primary objective of a user is to minimize the expected response time for their jobs. Since the response time at each server depends on the combined load from all users, a user's decision affects the performance others experience. In such a setting, no single user controls the entire system, and there is no communication among users, leading to inherently strategic behavior.

It is therefore natural to model the interaction among users as a non-cooperative game, where each user is a rational agent aiming to optimize its payoff (response time), and the strategic choices of all users determine the outcome. This game-theoretic formulation allows us to formally define the notion of equilibrium behavior, analyze the individual best responses and the conditions for equilibrium, and compare them with socially optimal solutions obtained via centralized control.
 We now define the non-cooperative resource allocation game.

\begin{defn}[Non-cooperative resource allocation game]
	\label{def:game}
	A non-cooperative resource allocation game is formally
	defined as the tuple
	$\Gamma=\langle [n],(\mathcal{S}_i)_{i\in[n]},
	(\mathcal{R}_i)_{i\in[n]}\rangle$,
	where $[n]$ is the set of players,
	$\mathcal{S}_i$ is the strategy set of player $i\in[n]$,
	and $\mathcal{R}_i$ is its expected response-time cost
	defined in \eqref{eq:drt}. Each player $i$ chooses a
	strategy $\mathbf{x}_i\in\mathcal{S}_i$, where
	\[
	\mathcal{S}_i=
	\left\{
	\mathbf{x}_i\in[0,1]^m:
	\sum_{j=1}^{m}x_{ij}=1
	\right\}.
	\]
	
The admissible strategies of player $i$ also
depend on the strategies of the other players
through the server-capacity constraints.

\end{defn}

In $\Gamma$, each player $i\in[n]$ seeks to minimize
its expected response time by solving
$\mathsf{DISOPTIM}_i$, given the strategies of
the other players.

\begin{defn}[Nash equilibrium]\label{def:nash}
	A feasible strategy profile
	$\mathbf{X}^*=(\mathbf{x}_i^*,\mathbf{x}_{-i}^*)$
	is a Nash equilibrium (NE) of $\Gamma$ if, for every
	$i\in[n]$,
	\[
	\mathbf{x}_i^*\in
	\argmin_{\mathbf{x}_i\in
		\mathcal{S}_i(\mathbf{x}_{-i}^*)}
	\mathcal{R}_i(\mathbf{x}_i,\mathbf{x}_{-i}^*),
	\]
	where $\mathcal{S}_i(\mathbf{x}_{-i}^*)$
	denotes the set of strategies of player $i$
	satisfying constraints
	\eqref{eq:con1}--\eqref{eq:con3},
	given the fixed strategies $\mathbf{x}_{-i}^*$
	of the other players.
\end{defn}

\subsection{Characterizing the Nash Equilibrium}
We first establish the unique best response of a user for fixed
strategies of the other users. An NE, in contrast,
requires the best-response conditions to hold simultaneously for
all users. We therefore distinguish individual optimality from
the equilibrium conditions of the complete game.

\begin{lemma}\label{prop:1}
For fixed $\mathbf{x}_{-i}$, the feasible set
of $\mathsf{DISOPTIM}_i$, defined by
\eqref{eq:con1}--\eqref{eq:con3}, is convex.
\end{lemma}

\begin{proof}
For fixed $\mathbf{x}_{-i}$,
constraint~\eqref{eq:con1} is a linear equality,
whereas \eqref{eq:con2} and \eqref{eq:con3}
are linear inequalities. Hence, the feasible
set is convex.
\end{proof}

\begin{lemma}\label{prop:2}
	The objective function~\eqref{eq:drt} is strictly convex over the feasible region.
\end{lemma}

\begin{proof}
	The function $\mathcal{R}_i(\mathbf{x}_i, \mathbf{x}_{-i})$ is additively separable as $\sum_j g_{ij}(x_{ij})$ with
	$
	g_{ij}(x_{ij}) = \frac{x_{ij}}{\mu_j^i - x_{ij} \phi_i}.
	$ It is to easy to verify that $g_{ij}$ is twice differentiable on the domain $x_{ij} \in [0, \mu_j^i/\phi_i)$, with
	\[
	g'_{ij}(x_{ij}) = \frac{\mu_j^i}{(\mu_j^i - x_{ij} \phi_i)^2} > 0, \quad
	g''_{ij}(x_{ij}) = \frac{2 \mu_j^i \phi_i}{(\mu_j^i - x_{ij} \phi_i)^3} > 0,
	\]
	for all $x_{ij}$ such that $\mu_j^i > x_{ij} \phi_i$. Hence, each $g_{ij}$ is strictly convex, and so is their sum. 
\end{proof}
From Lemma~\ref{prop:1} and Lemma~\ref{prop:2}, we derive the following theorem, which establishes the existence and uniqueness of the optimal solution to the $\mathsf{DISOPTIM}_i$.
%%%%%%%%%%%%%%%%%%%%%%%

\begin{thm}\label{thm:existence-uniqueness}
	Fix $\mathbf{x}_{-i}$ and suppose that $\phi_i>0$,
	$\mu_j^i>0$ for every $j\in[m]$, and the feasible
	set defined by \eqref{eq:con1}--\eqref{eq:con3}
	is nonempty. Then $\mathsf{DISOPTIM}_i$ admits
	a unique optimal solution.
\end{thm}

\begin{proof}
	Let $\mathbf{x}_i'$ be a feasible strategy and
	$\Omega=\mathcal{R}_i(\mathbf{x}_i',\mathbf{x}_{-i})$.
	Let us consider another feasible strategy $\mathbf{x}_i$ such that
	$\mathcal{R}_i(\mathbf{x}_i,\mathbf{x}_{-i})\leq\Omega$.
	Since every term in \eqref{eq:drt} is nonnegative,
	for every $j\in[m]$, we have $\frac{x_{ij}}{\mu_j^i-\phi_i x_{ij}}
			\leq\Omega$, which implies that $
			x_{ij}\leq
			\frac{\Omega\mu_j^i}{1+\Omega\phi_i}.$

	Now, consider the set
	\[
	Q_i=\left\{
	\mathbf{x}_i\in[0,1]^m:
	\sum_{j=1}^m x_{ij}=1,\quad
	x_{ij}\leq
	\frac{\Omega\mu_j^i}{1+\Omega\phi_i},
	\ \forall j\in[m]
	\right\}.
	\]
	Since $\mathbf{x}_i'$ is feasible and
	$\mathcal{R}_i(\mathbf{x}_i',\mathbf{x}_{-i})
	=\Omega$, the above inequality also holds for
	$\mathbf{x}_i'$. Hence, $\mathbf{x}_i'\in Q_i$,
	and $Q_i$ is nonempty. Similarly, every feasible
	strategy $\mathbf{x}_i$ satisfying
	$\mathcal{R}_i(\mathbf{x}_i,\mathbf{x}_{-i})
	\leq\Omega$ belongs to $Q_i$.
	
	The set $Q_i$ is closed and bounded, and hence
	compact. Moreover, for every $\mathbf{x}_i\in Q_i$,
	$
	\mu_j^i-\phi_i x_{ij}
	\geq
	\frac{\mu_j^i}{1+\Omega\phi_i}>0.
	$
	Thus, $Q_i$ lies within the feasible region,
	and $\mathcal{R}_i$ is continuous on $Q_i$.
	Consequently, it attains a minimum at some
	$\mathbf{x}_i^*\in Q_i$.
	
	Since every feasible strategy outside $Q_i$
	has an objective value greater than $\Omega$,
	whereas
	$\mathcal{R}_i(\mathbf{x}_i,\mathbf{x}_{-i})
	\leq\Omega$, the strategy $\mathbf{x}_i^*$
	is a global minimizer of $\mathsf{DISOPTIM}_i$. Finally, by Lemma~\ref{prop:2},
	$\mathcal{R}_i$ is strictly convex over the
	convex feasible set established in
	Lemma~\ref{prop:1}. Therefore, the global
	minimizer is unique.
\end{proof}

Since a NE is a strategy profile where all players' best responses to others' strategies coincide, we present the characterization of a player's best response.

\begin{thm}\label{thm:dispopt_main}
	Assume that $\mu_1^i \geq \mu_2^i \geq \cdots \geq \mu_m^i$. Then the best response of player $i$ (i.e., the solution to $\mathsf{DISOPTIM}_i$) is given by
	\begin{align}\label{eq:main_dist}
		x_{ij}=\frac{1}{\phi_i}
		\left[\mu_j^i-\sqrt{\frac{\mu_j^i}{\alpha_i^d}}\right]_+,
		\qquad
		\sum_{j=1}^m x_{ij}=1,
	\end{align}
	where $\alpha_i^d>0$ is uniquely determined by the
	normalization and $[z]_+=\max\{z,0\}$. A server $j$ is
	used by player $i$ precisely when
	$\alpha_i^d>1/\mu_j^i$.

%	\begin{align} \label{eq:main_dist}
%		x_{ij} = 
%		\begin{cases}
%			\dfrac{1}{\phi_i} \bigg( \mu_j^i - \dfrac{ \sqrt{\mu_j^i} ( \sum\limits_{r=1}^{k} \mu_r^i - \phi_i ) }{ \sum\limits_{r=1}^{k} \sqrt{\mu_r^i} } \bigg),  \forall j \in [1, k], \\ 
%			0,  \qquad \qquad \qquad\qquad \forall j \in [k+1, m],
%		\end{cases}
%			\text{where, }	k = \argmax_{\ell \in [1,m]} \bigg( \mu_\ell^i < \bigg( \frac{ \sum\limits_{r=1}^{\ell} \mu_r^i - \phi_i }{ \sum\limits_{r=1}^{\ell} \sqrt{\mu_r^i} } \bigg)^2 \bigg).
%		\end{align}
\end{thm}

\begin{proof}
	
Since the objective function $\mathsf{DISOPTIM}_i$ is strictly convex over a convex set, the corresponding Lagrangian is
\begin{equation}\label{thm2:eq1}
	\mathcal{L}(\mathbf{x}_i,\alpha^d,\boldsymbol{\beta},\boldsymbol{\theta})
	~=~ \sum_{j=1}^{m}\frac{x_{ij}}{\mu_j^i - x_{ij}\phi_i}
	~-~\alpha^d\!\left(\sum_{j=1}^{m}x_{ij}-1\right)
	~-~\sum_{j=1}^{m}\beta_j x_{ij}
	~+~\sum_{j=1}^{m}\theta_j\big(x_{ij}\phi_i - \mu_j^i\big),
	\end{equation}
	where $\alpha^d\in\mathbb{R}$ is the multiplier for the equality constraint $\sum_{j}x_{ij}=1$, each $\beta_j\ge 0$ is the multiplier for the nonnegativity constraint $x_{ij}\ge0$, and each $\theta_j\ge 0$ is the multiplier for the feasibility constraint $x_{ij}\phi_i \le \mu_j^i$. By KKT conditions \cite{Boyd:2004:CO:993483}, the optimal solution $\mathbf{x}_i = (x_{i1}, \ldots, x_{im})$ satisfies:
	
	\begin{align}\label{thm2:eq2}
		\frac{\mu_j^i}{(\mu_j^i-\phi_i x_{ij})^2}
		-\alpha_i^d-\beta_j&=0,\qquad j\in[m],\\
		\sum_{j=1}^m x_{ij}&=1,\\
		\beta_jx_{ij}&=0,\quad \beta_j\geq0,\quad x_{ij}\geq0.
	\end{align}

%	\begin{align}\label{thm2:eq2}
%		\begin{cases}
%		&\frac{\mu_j^i}{(\mu_j^i - x_{ij}\phi_i)^2} - \alpha^d - \beta_j = 0, \quad \forall j, \\
%		&\sum_{j=1}^{m} x_{ij} = 1, \\
%		&\beta_j x_{ij} = 0, \quad \beta_j \geq 0, \quad \theta_j(x_{ij}\phi_i - \mu_j^i) = 0.
%		\end{cases}
%			\end{align}
	
	Since $\mu_j^i > x_{ij}\phi_i$ must hold for feasibility, we have $\theta_j = 0$. This leads to:
	\begin{align}\label{thm2:eq3}
		\frac{\mu_j^i}{(\mu_j^i - x_{ij}\phi_i)^2} \geq \alpha_i^d, \quad \forall j,
	\end{align}
	with equality when $x_{ij} > 0$. Solving \eqref{thm2:eq3} we get:
	\begin{align} \label{thm2:eq4}
		x_{ij} = 
		\begin{cases}
			\frac{1}{\phi_i} \left( \mu_j^i - \frac{ \sqrt{\mu_j^i} }{ \sqrt{\alpha_i^d} } \right),&  \text{when } \alpha_i^d > \frac{1}{\mu_j^i},\\
			0, & \text{otherwise}.
		\end{cases}
	\end{align}
%%%%%%%%%%%%%%%%%%
	So, only those $j$ for which $\alpha_i^d > \frac{1}{\mu_j^i}$ will have $x_{ij} > 0$. Therefore, there exists some index $k<m$ such that
	%%%%%%%%%%%%%%
	\begin{equation}\label{thm2:eq5}
		\frac{1}{\mu_k^i} < \alpha_i^d < \frac{1}{\mu_{k+1}^i}.
	\end{equation}
%%%%%%%%%%%%%%
	Recalling and using the constraint $\sum_{j=1}^{k} x_{ij} = 1$ in \eqref{thm2:eq4}, we get
	%%%%%%%%%%%%%%%%%%%%%%%
	\begin{equation}\label{thm2:eq6}
\sum_{j=1}^{k} \frac{1}{\phi_i} \left( \mu_j^i - \frac{ \sqrt{\mu_j^i} }{ \sqrt{\alpha_i^d} } \right) = 1,
	\end{equation}
%%%%%%%%%%%%%%	
which implies
%%%%%%%%%%%%%%%%%%%%%
	\begin{equation}\label{thm2:eq7}
		\alpha_i^d = \left( \frac{ \sum_{j=1}^{k} \sqrt{\mu_j^i} }{ \sum_{j=1}^{k} \mu_j^i - \phi_i } \right)^2.
	\end{equation}
%%%%%%%%%%%%%%%%%%%%%%%

Since $\sum_{j=1}^m x_{ij}=1$, the positive multiplier
$\alpha_i^d$ is the unique solution of
\[
\sum_{j=1}^m
\left[\mu_j^i-\sqrt{\mu_j^i/\alpha_i^d}\right]_+
=\phi_i.
\]
For any trial active prefix $\{1,\ldots,k_i\}$, put
\[
t_{k_i}=
\frac{\sum_{r=1}^{k_i}\mu_r^i-\phi_i}
{\sum_{r=1}^{k_i}\sqrt{\mu_r^i}}.
\]
The prefix is valid if
$\sqrt{\mu_{k_i}^i}>t_{k_i}$ and either $k_i=m$
or $t_{k_i}\geq\sqrt{\mu_{k_i+1}^i}$.
Then $1/\sqrt{\alpha_i^d}=t_{k_i}$, which gives
\eqref{eq:main_dist}.

%
%
%Now, from \eqref{thm2:eq5} and \eqref{thm2:eq7} together, we get the index $k$ that is the largest value such that the inequality holds, i.e.,
%	\begin{equation} \label{thm2:eq8}
%	k = \argmax_{\ell \in [1,n]} \left\{ \mu_\ell^i < \left( \frac{ \sum_{r=1}^{\ell} \mu_r^i - \phi_i }{ \sum_{r=1}^{\ell} \sqrt{\mu_r^i} } \right)^2 \right\}.
%\end{equation}
%Substituting \eqref{thm2:eq8} into the \eqref{thm2:eq4}, we get 
%	\begin{align}\label{thm2:eq9}
%		x_{ij} = 
%		\begin{cases}
%			\dfrac{1}{\phi_i} \left( \mu_j^i - \dfrac{ \sqrt{\mu_j^i} \Big( \sum\limits_{r=1}^{k} \mu_r^i - \phi_i \Big) }{ \sum\limits_{r=1}^{k} \sqrt{\mu_r^i} } \right), & \forall j \in [1, k], \\ 
%			0, & \forall j \in [k+1, m],
%		\end{cases}
%	\end{align}
%which is the desired result. 
%	
\end{proof}
\begin{remark}
Operationally, Theorem \ref{thm:dispopt_main} implies that player $i$ distributes its load only across the top $k$ most capable options (highest $\mu_j^i$), ignoring the rest. Within this selected set, the load is not divided equally; instead, it is allocated in a manner that reflects the relative speeds of the available options. Faster options, characterized by larger service rates $\mu_j^i$, are assigned a proportionally greater load share.
\end{remark}

%%%%%%%%%%%%%%%%%%%%%%%%%

\begin{fact}\label{fact:1}
	Let $\mathbf{X}^*=(\mathbf{x}_1^*,\ldots,\mathbf{x}_n^*)$
	be a feasible NE, and define
	$\lambda_j^d=\sum_{i=1}^n\phi_i x_{ij}^*$.
	For each $i\in[n]$ there exists $\alpha_i^d>0$ such that
	\begin{equation}\label{eq:fact11}
		\frac{1}{\mu_j-\lambda_j^d}
		+\frac{\phi_i x_{ij}^*}{(\mu_j-\lambda_j^d)^2}
		\begin{cases}
			=\alpha_i^d,&x_{ij}^*>0,\\
			\geq\alpha_i^d,&x_{ij}^*=0,
		\end{cases}
		\qquad \forall j\in[m].
	\end{equation}
	Moreover,
	\begin{equation}\label{eq:fact12}
		\sum_{j=1}^m x_{ij}^*=1\quad (\forall i\in[n]),
		\qquad
		\sum_{j=1}^m\lambda_j^d=\Phi,
		\qquad \lambda_j^d<\mu_j\quad(\forall j\in[m]).
	\end{equation}
	Conversely, any feasible profile satisfying these conditions
	for all $i\in[n]$ is an NE.
\end{fact}
\begin{proof}
	Fix $i$ and the equilibrium strategies of all other users.
	By the strict convexity of $\mathcal{R}_i$ in
	$\mathbf{x}_i$, the KKT conditions are
	necessary and sufficient for a best response. Differentiating
	\eqref{eq:drt} with respect to $x_{ij}$ yields
	\[
	\frac{\partial\mathcal{R}_i}{\partial x_{ij}}
	=\frac{1}{\mu_j-\lambda_j^d}
	+\frac{\phi_i x_{ij}}{(\mu_j-\lambda_j^d)^2}.
	\]
	Complementary slackness gives \eqref{eq:fact11}; feasibility
	gives \eqref{eq:fact12}. The converse follows by applying
	the same sufficient conditions to every user.
\end{proof}

\begin{thm} 
	\label{thm:symmetric-ne}
	Suppose $n\geq2$ and $\phi_i=\Phi/n$ for every $i\in[n]$,
	where $0<\Phi<\sum_{j=1}^m\mu_j$. For $\alpha>0$, define
	\[
	d_j(\alpha)=
	\frac{n-1+\sqrt{(n-1)^2+4n\alpha\mu_j}}{2n\alpha},
	\qquad
	\lambda_j(\alpha)=\bigl[\mu_j-d_j(\alpha)\bigr]_+.
	\]
 
There is a unique $\alpha>0$
satisfying $\sum_{j=1}^m\lambda_j(\alpha)=\Phi$.
The strategy profile
$
x_{ij}^*=\frac{\lambda_j(\alpha)}{\Phi}, i\in[n],\ j\in[m],
$
is the unique symmetric NE.  
\end{thm}
\begin{proof}
	At the stated profile, the load of player $i$ on server $j$
	is $\phi_i x_{ij}=\lambda_j/n$. Setting the marginal
	response time in \eqref{eq:fact11} equal to $\alpha$ on
	active servers gives
	\[
	\frac{1}{d_j}+\frac{\lambda_j/n}{d_j^2}=\alpha,
	\qquad d_j=\mu_j-\lambda_j.
	\]
	Consequently,
	$n\alpha d_j^2-(n-1)d_j-\mu_j=0$, whose positive root is
	$d_j(\alpha)$. A server is inactive exactly when
	$\alpha\leq1/\mu_j$, consistent with the positive-part
	formula. Each $\lambda_j(\alpha)$ is continuous and
	nondecreasing in $\alpha$, strictly increasing after it
	becomes positive. The sum increases continuously from zero
	to $\sum_j\mu_j$; hence a unique $\alpha$ achieves $\Phi$. Thus, every player satisfies its best-response
conditions, which are sufficient by strict
convexity. The uniqueness of $\alpha$ also
ensures the uniqueness of the symmetric NE.
\end{proof}

%%%%%%%%%%%%%%%%%%%%%%%%%%
%\begin{fact}\label{fact:1}
%	Let $\lambda_j^{d}$ denote the total load assigned to server $j$ under the optimal strategy given in Theorem \ref{thm:dispopt_main} (distributed allocation strategy): $\lambda_j^d = \sum_{i=1}^{n} x_{ij} \phi_i$. Recalling the available processing rate on server $j$ for any player $i$, denoted by $\mu_{j}^{i}$ from \eqref{eq:availcap}, we have that $
%		\mu_{j}^{i} \leq \mu_{j}$. Now, From \eqref{thm2:eq4} and \eqref{eq:availcap}, the load $\lambda_j^{d}$ can be expressed as a function of $\alpha^{d}$:
%%%%%%%%%%%%%%%	
%	\begin{empheq}[left=\empheqlbrace]{align}
%		&\lambda_j^d (\alpha^d) = \mu_j - \frac{\sqrt{\mu_j^i}}{\sqrt{\alpha^{d}}}, \quad \alpha^d > \frac{1}{\mu_j^i} \label{eq:fact11} \\
%		&\sum_{j=1}^{k} \lambda_j^d (\alpha^d) = \Phi \label{eq:fact12}
%	\end{empheq}
%%%%%%%%%%%%%%%%%
%\end{fact}
%%%%%%%%%%%%%%%%%%%%%%%%%
\subsection{Characterizing the Optimal Strategy for Centralized Model}

In the centralized model, the centralized dispatcher determines the complete allocation vector by jointly optimizing over the strategies of all users subject to the global constraints, given in \eqref{eq:ccon1}-- \eqref{eq:ccon3}. We now characterize the socially optimal strategy profile in closed form by solving the global problem \textsf{SOCOPTIM}.

For convenience, we define
\[
a_k=\sum_{j=1}^{k}\sqrt{\mu_j},
\qquad
b_k=\sum_{j=1}^{k}\mu_j,
\quad k\in[m].
\]

\begin{thm}\label{thm:thmsoc}
	Let $\mu_1\geq\cdots\geq\mu_m>0$ and
	$0<\Phi<\sum_{j=1}^m\mu_j$. There exists a unique
	$t>0$ satisfying
	\[
	\sum_{j=1}^m[\mu_j-t\sqrt{\mu_j}]_+=\Phi.
	\]
	The socially optimal routing fractions are
	\begin{equation}\label{eq:mainsoc}
		s_j=\frac{[\mu_j-t\sqrt{\mu_j}]_+}{\Phi},
		\qquad j\in[m].
	\end{equation}
	If the active set is $\{1,\ldots,k_s\}$, then
	$
	t=\frac{b_{k_s}-\Phi}{a_{k_s}}$,
$\sqrt{\mu_{k_s}}>t$,
	$ k_s=m\ \text{or}\ t\geq\sqrt{\mu_{k_s+1}}.
	$
\end{thm}

\begin{proof}
	Following the proof of Theorem~\ref{thm:dispopt_main},
	we apply the KKT conditions to $\mathsf{SOCOPTIM}$.
	Let $\alpha^s$ be the Lagrange multiplier associated
	with $\sum_{j=1}^{m}s_j=1$. For every active server
	$j$, the stationarity condition gives
	\begin{flalign*}
		&\frac{\mu_j}{(\mu_j-s_j\Phi)^2}=\alpha^s
		&&\\
		\implies&
		s_j=\frac{1}{\Phi}
		\left(\mu_j-\frac{\sqrt{\mu_j}}{\sqrt{\alpha^s}}\right).
		&&
	\end{flalign*}
	
	For an inactive server, the KKT conditions require
	$\alpha^s\leq1/\mu_j$. Setting
	$t=1/\sqrt{\alpha^s}$, both cases yield
	\begin{flalign*}
		& s_j=\frac{[\mu_j-t\sqrt{\mu_j}]_+}{\Phi}.
		&&
	\end{flalign*}
	
	Using $\sum_{j=1}^{m}s_j=1$, we obtain
	\begin{flalign*}
		&\sum_{j=1}^{m}
		[\mu_j-t\sqrt{\mu_j}]_+=\Phi.
		&&
	\end{flalign*}
	The left-hand side is continuous and strictly
	decreasing from $\sum_{j=1}^{m}\mu_j$ to $0$
	as $t$ increases from $0$ to $\sqrt{\mu_1}$.
	Since $0<\Phi<\sum_{j=1}^{m}\mu_j$, there
	exists a unique $t>0$ satisfying this equation.
	
	Since $\mu_1\geq\cdots\geq\mu_m$, the active
	servers form the set $\{1,\ldots,k_s\}$. Thus,
	\begin{flalign*}
		&\sum_{j=1}^{k_s}
		(\mu_j-t\sqrt{\mu_j})=\Phi
		&&\\
		\implies&
		t=\frac{b_{k_s}-\Phi}{a_{k_s}}.
		&&
	\end{flalign*}
	The active set conditions are
	$\sqrt{\mu_{k_s}}>t$ and either $k_s=m$
	or $t\geq\sqrt{\mu_{k_s+1}}$.
	
	The resulting allocation satisfies all feasibility
	and KKT conditions. Since $\mathcal{C}(\mathbf{s})$
	is strictly convex over the feasible region,
	the allocation is the unique global optimum.
\end{proof}

 From Theorem~\ref{thm:thmsoc}, the total load
 assigned to server $j$ under the socially optimal
 allocation is $\lambda_j^s=s_j\Phi$. Setting
 $t=1/\sqrt{\alpha^s}$, we obtain
 \begin{flalign}
 	&\lambda_j^s(\alpha^s)
 	=\left[\mu_j-
 	\frac{\sqrt{\mu_j}}{\sqrt{\alpha^s}}\right]_+,
 	\quad j\in[m], \label{eq:fact21} &&\\
 	&\sum_{j=1}^{m}\lambda_j^s(\alpha^s)
 	=\Phi. \label{eq:fact22} &&
 \end{flalign}

 \section{Inefficiency of Distributed Allocation: Tragedy of the Commons}
 \label{sec:esttoc}
 
 We now compare the load distribution at a Nash equilibrium with the
 socially optimal allocation. In particular, we examine which servers
 are used and how the total load is distributed among them. The
 comparison must use the simultaneous equilibrium conditions in
 Fact~\ref{fact:1}, because the individual multipliers $\alpha_i^d$
 need not equal the centralized multiplier $\alpha^s$.
 
 Let $\mathbf{X}^*$ be a feasible Nash equilibrium and write
 $\lambda_j^d=\sum_{i=1}^n\phi_i x_{ij}^*$ and
 $\lambda_j^s=s_j\Phi$. Assume throughout that
 $\mu_1\geq\mu_2\geq\cdots\geq\mu_m>0$. Define
 $\mathcal{K}_d=\{j:\lambda_j^d>0\}$ and
 $\mathcal{K}_s=\{j:\lambda_j^s>0\}$, with respective
 cardinalities $k_d$ and $k_s$. Also let
 $d_j=\mu_j-\lambda_j^d>0$ denote the residual capacity
 of server $j$ at equilibrium.
 
 \begin{lemma}\label{claim:ineffi1}
 	The residual capacities satisfy
 	$d_1\geq d_2\geq\cdots\geq d_m$.
 	Consequently, the servers used at Nash equilibrium form
 	an initial segment $\mathcal{K}_d=\{1,\ldots,k_d\}$.
 \end{lemma}
 
 \begin{proof}
 	By \eqref{eq:fact11}, for every
 	$i\in[n]$ and $j\in[m]$, we have
 	\begin{equation}\label{eq:lemm3_eq1}
 		\phi_i x_{ij}^*
 		=d_j[\alpha_i^d d_j-1]_+.
 	\end{equation}
  Summing \eqref{eq:lemm3_eq1} over all users and using $\lambda_j^d=\mu_j-d_j$, we obtain 
 \begin{equation}\label{eq:dist-residual}
 	\mu_j=d_j+\sum_{i=1}^n
 	d_j[\alpha_i^d d_j-1]_+.
 \end{equation}
 Since each term $d_j[\alpha_i^d d_j-1]_+$
 is nondecreasing in $d_j>0$, the right-hand
 side in \eqref{eq:dist-residual} is strictly increasing in $d_j$.
 	  Hence,
 	$\mu_1\geq\cdots\geq\mu_m$ implies
 	$d_1\geq\cdots\geq d_m$. Moreover,
 	$\lambda_j^d>0$ if and only if
 	$\alpha_i^d d_j>1$ for at least one user $i$.
 	Thus, if server $j$ is used, every server with a
 	smaller index is also used.
 \end{proof}
 
 \begin{lemma}\label{claim:2}
 	For the servers used at NE, the ratios
 	$
 	\gamma_j=\frac{d_j}{\sqrt{\mu_j}}
 $
 	satisfy $\gamma_1\leq\gamma_2\leq\cdots\leq\gamma_{k_d}$.
 \end{lemma}

 \begin{proof}
 	From \eqref{eq:dist-residual}, we have
 	\begin{equation}\label{eq:lemma4_eq1}
 		\frac{\mu_j}{d_j^2}
 		=\frac{1}{d_j}
 		+\sum_{i=1}^{n}
 		\left[\alpha_i^d-\frac{1}{d_j}\right]_+.
 	\end{equation}
 	
 	Consider any two servers $j$ and $\ell$, such that
 	$j<\ell\leq k_d$, which are active at NE.
 	By Lemma~\ref{claim:ineffi1}, $d_j\geq d_\ell$.
 	Since server $\ell$ is used at NE, there exists
 	at least one player $i$ such that
 	$\alpha_i^d d_\ell>1$. 	Now, using \eqref{eq:lemma4_eq1}, we obtain
 	\begin{equation}
 		\frac{\mu_j}{d_j^2}
 		-\frac{\mu_\ell}{d_\ell^2}
 		=\frac{1}{d_j}-\frac{1}{d_\ell}
 		+\sum_{i=1}^{n}
 		\left(
 		\left[\alpha_i^d-\frac{1}{d_j}\right]_+
 		-
 		\left[\alpha_i^d-\frac{1}{d_\ell}\right]_+
 		\right).
 	\end{equation}
 	Moreover, for at least one player
 		\[
 	\left[\alpha_i^d-\frac{1}{d_j}\right]_+
 	-
 	\left[\alpha_i^d-\frac{1}{d_\ell}\right]_+
 	=
 	\frac{1}{d_\ell}-\frac{1}{d_j},
 	\]
 	 while the remaining
 	differences are nonnegative. Therefore,
 	\[
 	\frac{\mu_j}{d_j^2}
 	-\frac{\mu_\ell}{d_\ell^2}\geq0,
 	\]
 	which implies
 	\[
 	\frac{d_j}{\sqrt{\mu_j}}
 	\leq
 	\frac{d_\ell}{\sqrt{\mu_\ell}}.
 	\]
 	Hence,
 	$\gamma_1\leq\gamma_2\leq\cdots
 	\leq\gamma_{k_d}$.
 \end{proof}

 \begin{lemma}\label{lemma:3}
 	Every server used at an NE is also used
 	at the social optimum. In particular,
 	$\mathcal{K}_d\subseteq\mathcal{K}_s$ and $k_d\leq k_s$.
 \end{lemma}

 \begin{proof}
 	Let $t=\frac{\sum_{j=1}^{k_s}\mu_j-\Phi}
 	{\sum_{j=1}^{k_s}\sqrt{\mu_j}},$ as defined in Theorem~\ref{thm:thmsoc},
 	so that $\lambda_j^s=[\mu_j-t\sqrt{\mu_j}]_+$. We prove the lemma by contradiction. Assume that $k_d>k_s$. Server $k_s+1$ is used at NE but not at
 	the social optimum. 
 	Consequently, $\sqrt{\mu_{k_s+1}}\leq t$.
 	Moreover, since server $k_s+1$ is active at NE,
 	$\lambda_{k_s+1}^d>0$, which implies
 	$
 	d_{k_s+1}
 	=\mu_{k_s+1}-\lambda_{k_s+1}^d
 	<\mu_{k_s+1}.
 	$
 	 Therefore,
 	$
 	\gamma_{k_s+1}
 	=\frac{d_{k_s+1}}{\sqrt{\mu_{k_s+1}}}
 	<\sqrt{\mu_{k_s+1}}\leq t$. By Lemma~\ref{claim:2},
 	$\gamma_j\leq\gamma_{k_s+1}<t$
 	for every $j\leq k_s$. Hence, for $j\leq k_s$
 	\begin{flalign*}
 		\lambda_j^d
 		=&~\mu_j-\gamma_j\sqrt{\mu_j}&&\\
 		>& ~\mu_j-t\sqrt{\mu_j}&&\\
 		=&~\lambda_j^s&&.
 		\end{flalign*}
 	
 	Moreover,
 	$\lambda_{k_s+1}^d>0$ and $\lambda_{k_s+1}^s=0$.
 	Thus, the total load at NE exceeds that
 	of the social optimum, contradicting
 	\[
 	\sum_{j=1}^m\lambda_j^d
 	=\sum_{j=1}^m\lambda_j^s=\Phi.
 	\]
 	Therefore, $k_d\leq k_s$, which implies
 	$\mathcal{K}_d\subseteq\mathcal{K}_s$.
 \end{proof}

 The above lemmas establish the structure of the active
 sets and make it possible to compare the two actual load
 allocations without identifying their Lagrange multipliers.
 
	 \begin{thm} \label{thm:main_inefficiency}
	 	There exists an index $\ell\in\{1,\ldots,k_s+1\}$ such that
	 	\begin{enumerate}[label=(\roman*)]
	 		\item \label{item:1thm_main}
	 		$\lambda_j^d\geq\lambda_j^s$ for every $j<\ell$;
	 		\item \label{item:2thm_main}
	 		$\lambda_j^d\leq\lambda_j^s$ for every $j\geq\ell$.
	 	\end{enumerate}
	 	Furthermore, for every $r\in[m]$,
	 	\begin{equation}\label{eq:prefix-load-shift}
	 		\sum_{j=1}^r\lambda_j^d
	 		\geq\sum_{j=1}^r\lambda_j^s.
	 	\end{equation}
	 	 
	 \end{thm}

 \begin{proof}
 	Let
 	$t=\frac{\sum_{j=1}^{k_s}\mu_j-\Phi}
 	{\sum_{j=1}^{k_s}\sqrt{\mu_j}}$,
 	as defined in Theorem~\ref{thm:thmsoc}.
 	By Lemma~\ref{lemma:3}, $k_d\leq k_s$.
 	For every $j\leq k_d$, we have
 	\begin{flalign*}
 		\lambda_j^d-\lambda_j^s
 		=&~(\mu_j-\gamma_j\sqrt{\mu_j})
 		-(\mu_j-t\sqrt{\mu_j})&&\\
 		=&~(t-\gamma_j)\sqrt{\mu_j}&&.
 	\end{flalign*}
 	
 	\noindent \textit{Part (i):}
 	By Lemma~\ref{claim:2},
 	$\gamma_1\leq\gamma_2\leq\cdots
 	\leq\gamma_{k_d}$.
 	Let $\ell$ be the smallest index such that
 	$\gamma_\ell\geq t$. If no such index exists,
 	let $\ell=k_d+1$.
 	For every $j<\ell$, we have $\gamma_j<t$.
 	Therefore,
 	\[
 	\lambda_j^d-\lambda_j^s
 	=(t-\gamma_j)\sqrt{\mu_j}>0,
 	\]
 	which implies $\lambda_j^d\geq\lambda_j^s$
 	for every $j<\ell$.
 	
 	\noindent\textit{Part (ii):}
 	For every $\ell\leq j\leq k_d$,
 	Lemma~\ref{claim:2} gives $\gamma_j\geq t$.
 	Consequently,
 	\[
 	\lambda_j^d-\lambda_j^s
 	=(t-\gamma_j)\sqrt{\mu_j}\leq0.
 	\]
 	Moreover, for every $j>k_d$, we have
 	$\lambda_j^d=0\leq\lambda_j^s$.
 	Hence, $\lambda_j^d\leq\lambda_j^s$
 	for every $j\geq\ell$.
 	
 	\noindent\textit{For \eqref{eq:prefix-load-shift}:}
 	For $r<\ell$, Part (i) directly gives
 	$
 	\sum_{j=1}^{r}\lambda_j^d
 	\geq
 	\sum_{j=1}^{r}\lambda_j^s.
 $
 	
 	For $r\geq\ell$, Part (ii) implies
 	\[
 	\sum_{j=r+1}^{m}\lambda_j^d
 	\leq
 	\sum_{j=r+1}^{m}\lambda_j^s.
 	\]
 	Since the total load under both allocations
 	is $\Phi$, we obtain
 	\begin{flalign*}
 		\sum_{j=1}^{r}\lambda_j^d
 		=&~\Phi-\sum_{j=r+1}^{m}\lambda_j^d\\
 		\geq&~\Phi-\sum_{j=r+1}^{m}\lambda_j^s\\
 		=&~\sum_{j=1}^{r}\lambda_j^s.
 	\end{flalign*}
 	Thus, \eqref{eq:prefix-load-shift} holds
 	for every $r\in[m]$.
 \end{proof}

\begin{remark}
	Theorem~\ref{thm:main_inefficiency} establishes the structural origin of the ToC in distributed resource allocation. At NE, selfish users collectively assign excessive load to faster servers, whereas the social optimum redistributes part of this load toward slower servers to reduce overall congestion. This difference between the two allocations gives rise to the ToC in our setting. However, the crossing index $\ell$ need not equal $k_d+1$, since a server may receive less load at NE than at the social optimum even when it is active in both allocations. Thus, the ToC arises from the collective concentration of load on faster servers,
	rather than the overloading of every active server.
\end{remark}

 \section{Quantifying Inefficiency: Price of Anarchy}\label{sec:quantpoa}
 
 Theorem~\ref{thm:main_inefficiency} shows that, compared with the social
 optimum, the distributed equilibrium places more cumulative load on the
 faster servers. However, this result does not quantify the resulting
 loss in system efficiency. We therefore study the Price of Anarchy
 ($\PoA$), which measures the inefficiency caused by selfish resource
 allocation~\cite{roughgarden2002bad,christodoulou2005price}.
 
 For any feasible allocation $\pi$, the total system cost is
 \begin{equation}\label{eq:poacostmdl}
 	C(\pi)=\sum_{j=1}^{m}
 	\frac{\mathcal{W}_j(\pi)}{\mu_j-\mathcal{W}_j(\pi)},
 \end{equation}
 where $\mathcal{W}_j(\pi)$ denotes the aggregate load assigned to
 server $j$. In particular,
 $\mathcal{W}_j(\pi_d)=\lambda_j^d$ at an NE and
 $\mathcal{W}_j(\pi_s)=\lambda_j^s$ at the social optimum.
 Thus, the total system cost under the socially
 optimal allocation is $
 C(\pi_s)=\Phi\mathcal{C}(\mathbf{s})$.

 \begin{defn}[Price of Anarchy]
 	Let $\mathsf{NE}(\Gamma)$ denote the nonempty set of Nash equilibria
 	of $\Gamma$, and let $\pi_s$ denote the socially optimal allocation.
 	The PoA of an instance $ \mathscr{I}$ is
 	\[
 	\PoA(\mathscr{I})
 	=\sup_{\pi_d\in\mathsf{NE}(\Gamma)}
 	\frac{C(\pi_d)}{C(\pi_s)}.
 	\]
 	For a specified equilibrium $\pi_d$, the ratio
 	$C(\pi_d)/C(\pi_s)$ is its equilibrium cost ratio.
 \end{defn}
 
 Recall that the servers are ordered as
 $\mu_1\geq\cdots\geq\mu_m>0$. For $k\in[m]$, let
 \begin{equation}\label{eq:poa1}
 	a_k=\sum_{j=1}^{k}\sqrt{\mu_j},\qquad
 	b_k=\sum_{j=1}^{k}\mu_j.
 \end{equation}
 These quantities describe the social optimum through
 Theorem~\ref{thm:thmsoc}. The equilibrium cost, in contrast,
 depends on the actual equilibrium loads $\lambda_j^d$.
 
 \begin{thm}\label{thm:exactpoa}
 	Let $\pi_d$ be a Nash equilibrium with
 	$\lambda_j^d=\sum_{i=1}^n\phi_i x_{ij}^*$, and let
 	$\pi_s$ be the socially optimal allocation using the servers
 	$\{1,\ldots,k_s\}$. Then
 	\begin{equation}\label{eq:exactpoa-corrected}
 		\frac{C(\pi_d)}{C(\pi_s)}
 		=\frac{\displaystyle\sum_{j=1}^{m}
 			\frac{\lambda_j^d}{\mu_j-\lambda_j^d}}
 		{\displaystyle\frac{a_{k_s}^{2}}{b_{k_s}-\Phi}-k_s}.
 	\end{equation}
 	If the equilibrium is unique, this ratio equals
 	$\PoA( \mathscr{I})$; otherwise, the PoA is the supremum
 	of this expression over all Nash equilibria.
 \end{thm}
 
 \begin{proof}
 	Let $t=(b_{k_s}-\Phi)/a_{k_s}$, as in
 	Theorem~\ref{thm:thmsoc}. For $j\leq k_s$,
 	$\lambda_j^s=\mu_j-t\sqrt{\mu_j}$, and
 	$\lambda_j^s=0$ otherwise. Therefore,
 	\begin{flalign*}
 		C(\pi_s)
 		&=\sum_{j=1}^{k_s}
 		\frac{\mu_j-t\sqrt{\mu_j}}{t\sqrt{\mu_j}}
 		&&\\
 		&=\frac{a_{k_s}^{2}}{b_{k_s}-\Phi}-k_s. &&
 	\end{flalign*}
 	Moreover, by \eqref{eq:poacostmdl},
 	$C(\pi_d)=\sum_{j=1}^{m}
 	\lambda_j^d/(\mu_j-\lambda_j^d)$.
 	Dividing the two costs proves the result.
 \end{proof}
 
 \paragraph{Discussion on Theorem~\ref{thm:exactpoa}}
 The social cost is determined by the active social-optimum servers
 through $a_{k_s}$ and $b_{k_s}$. The distributed cost also depends
 on the actual server loads $\lambda_j^d$. Although the individual
 routing fractions $x_{ij}^*$ enter the cost only through their
 aggregate loads, the number of active servers alone does not determine
 the equilibrium cost. This distinction follows from the different
 optimality conditions in Fact~\ref{fact:1} and
 Theorem~\ref{thm:thmsoc}.
 
 \begin{corollary}\label{cor:poa-lower-bound}
 	For every feasible instance $ \mathscr{I}$ with at least one NE,
 	$\PoA( \mathscr{I})\geq1$.
 	Moreover, $\PoA( \mathscr{I})>1$ whenever some NE
 	has strictly greater system cost than the social optimum.
 \end{corollary}
 
 \begin{proof}
 	Since $\pi_s$ minimizes the total system cost over all feasible
 	allocations and every NE $\pi_d$ is feasible,
 	$C(\pi_s)\leq C(\pi_d)$. As $C(\pi_s)>0$, the definition of
 	$\PoA$ gives $\PoA( \mathscr{I})\geq1$, with strict inequality
 	if $C(\pi_d)>C(\pi_s)$ for some equilibrium.
 \end{proof}
 
 The following example illustrates the efficiency loss at a
 symmetric Nash equilibrium.
 
 \begin{exm} 
 	Let $n=2$, $m=2$, $\phi_1=\phi_2=4$,
 	$\mu_1=8.5$ and $\mu_2=0.6$, so that $\Phi=8$.
 	By Theorem~\ref{thm:symmetric-ne}, the symmetric NE satisfies
 	\[
 	\frac{1}{8.5-\lambda_1^d}
 	+\frac{\lambda_1^d/2}{(8.5-\lambda_1^d)^2}
 	=
 	\frac{1}{0.6-\lambda_2^d}
 	+\frac{\lambda_2^d/2}{(0.6-\lambda_2^d)^2},
 	\quad \lambda_1^d+\lambda_2^d=8.
 	\]
 	Solving gives
 	$(\lambda_1^d,\lambda_2^d)\approx(7.655521,0.344479)$,
 	with $C(\pi_d)\approx10.413521$.
 	Theorem~\ref{thm:thmsoc} gives
 	$(\lambda_1^s,\lambda_2^s)\approx(7.630905,0.369095)$,
 	with $C(\pi_s)\approx10.378760$. Hence,
 $
 	\frac{C(\pi_d)}{C(\pi_s)}\approx1.003349.
 $
 	This is the equilibrium cost ratio for the displayed symmetric NE. Thus, even in a simple two-server system, the
 	NE incurs approximately $0.335\%$ higher cost than the socially optimal
 	allocation. The reason for this may be attributed to the fact that selfish
 	users assign more load to the faster
 	server, increasing congestion relative to the
 	social optimum. Although the efficiency loss is
 	small in this instance, the example illustrates
 	the ToC caused by selfish resource allocation.
 \end{exm}

 \begin{exm} 
 	Consider $n=2$ users with $\phi_1=\phi_2=3.25$
 	and $m=2$ servers with $\mu_1=8.5$ and
 	$\mu_2=0.6$. Thus, $\Phi=6.5$.
 	
 	Suppose both users route their entire load to
 	server $1$, giving $(\lambda_1^d,\lambda_2^d)
 	=(6.5,0)$. By Fact~\ref{fact:1}, this profile
 	is a NE since
 	$\frac{\mu_1-\lambda_1^d+\phi_i x_{i1}}{(\mu_1-\lambda_1^d)^2}
 	=\frac{8.5-3.25}{(8.5-6.5)^2}
 	=1.3125<\frac{1}{0.6}
 	=\frac{1}{\mu_2-\lambda_2^d}$.
 	Hence, $k_d=1$ and
 	$
 	C(\pi_d)=\frac{6.5}{8.5-6.5}=3.25$.

 	 By Theorem~\ref{thm:thmsoc},
 	$
 	t=\frac{8.5+0.6-6.5}
 	{\sqrt{8.5}+\sqrt{0.6}}
 	\approx0.704593<\sqrt{0.6}.
 $
 	Thus, both servers are active at the social
 	optimum, i.e., $k_s=2$, with
 	$
 	(\lambda_1^s,\lambda_2^s)
 	\approx(6.445776,0.054224).
 $
 	Consequently,
 	$
 	C(\pi_s)\approx3.237168$, and $
 	\frac{C(\pi_d)}{C(\pi_s)}\approx1.003964.
 $
 	
 	Thus, the NE uses fewer servers than
 	the social optimum and incurs approximately
 	$0.396\%$ higher system cost. This illustrates
 	that selfish routing may leave a server unused
 	even when its utilization improves system efficiency.
 \end{exm}

 We next show that the PoA has no uniform upper bound across
 all instances. This requires a family of instances; it does not
 follow merely from $k_d<k_s$ for a fixed set of servers.
 
 \begin{thm}\label{thm:unboundedpoa}
 	The PoA of the distributed resource allocation game
 	is unbounded over the class of instances with arbitrary numbers
 	of users and heterogeneous servers.
 \end{thm}
 %%%%%

 \begin{proof}
 	For each integer $m\geq2$, consider a system with
 	$m$ servers having service rates
 	$\mu_1=1$ and $\mu_j=1/m^2$ for
 	$j=2,\ldots,m$.
 	
 	According to \eqref{eq:poa1},
 	we have
 	$
 		a_m
 		=~\sum_{j=1}^{m}\sqrt{\mu_j}
 		=2-\frac{1}{m}$, 
 		$b_m
 		=~\sum_{j=1}^{m}\mu_j
 		=1+\frac{m-1}{m^2}$. Let the total incoming load be
 	$\Phi=b_m-\frac{1}{2m}$, distributed equally
 	among $n$ users, i.e.,
 	$\phi_i=\Phi/n$ for every $i\in[n]$.
 	
 	From Theorem~\ref{thm:thmsoc}, we have
 $
 	t=\frac{b_m-\Phi}{a_m}
 	=\frac{1}{2ma_m}<\frac{1}{m}.
 $
 	Since $\sqrt{\mu_m}=1/m$, all $m$ servers
 	are active at the social optimum. Therefore,
 	\begin{equation}\label{eq:thm7_eq1}
 			C(\pi_s)=\frac{a_m^2}{b_m-\Phi}-m
 		=2ma_m^2-m.
 	\end{equation}
 By Theorem~\ref{thm:symmetric-ne}, a symmetric
NE exists for every $n$. From Fact~\ref{fact:1},
for each active server $j$,
$\alpha=\frac{1}{d_j}
+\frac{\lambda_j^d}{nd_j^2}$, with $d_j=\mu_j-\lambda_j^d.$ As $n\to\infty$, $d_j\to1/\alpha$ and hence
$\lambda_j^d\to\mu_j-1/\alpha$.
For inactive servers, Fact~\ref{fact:1}
gives $\alpha\leq1/\mu_j$ and their
loads remain zero. Thus,
$
\lambda_j^d\longrightarrow
\left[\mu_j-\frac{1}{\alpha}\right]_+.
$
Since $\sum_{j=1}^m\lambda_j^d=\Phi$,
we obtain
\begin{equation}\label{eq:thm7_eq_2}
	\sum_{j=1}^{m}
	\left[\mu_j-\frac{1}{\alpha}\right]_+
	=\Phi.
\end{equation}

Suppose all $m$ servers are active as $n\to\infty$.
Using \eqref{eq:thm7_eq_2}, we have
$\sum_{j=1}^{m}(\mu_j-1/\alpha)=\Phi$.
Since $\sum_{j=1}^{m}\mu_j=b_m$ and
$\Phi=b_m-1/(2m)$, we obtain
$b_m-m/\alpha=b_m-1/(2m)$, yielding
$\alpha=2m^2$.
Since $\mu_m=1/m^2$ is the smallest server
capacity, $\alpha>1/\mu_j$ holds for every
$j\in[m]$. This confirms our assumption
that all servers are active in the limiting
allocation.

Since all servers are active and
$d_j\to1/\alpha=1/(2m^2)$ as $n\to\infty$,
the equilibrium cost satisfies
\begin{flalign*}
	\lim_{n\to\infty}C(\pi_d)
	&=\lim_{n\to\infty}
	\sum_{j=1}^{m}\frac{\lambda_j^d}{d_j}&&\\
	&=\lim_{n\to\infty}
	\sum_{j=1}^{m}
	\left(\frac{\mu_j}{d_j}-1\right)&&\\
	&=2m^2b_m-m.&&
\end{flalign*}

Now, we obtain
\begin{flalign*}
	\lim_{n\to\infty}
	\frac{C(\pi_d)}{C(\pi_s)}
	&=\frac{2m^2b_m-m}{2ma_m^2-m}&&\\
	&=\frac{2mb_m-1}{2a_m^2-1}&&\\
	&=\frac{
		2m\left(1+\frac{m-1}{m^2}\right)-1
	}{
		2\left(2-\frac{1}{m}\right)^2-1
	}.&&
\end{flalign*}

Consequently,
\begin{flalign*}
	\lim_{m\to\infty}\lim_{n\to\infty}
	\frac{C(\pi_d)}{C(\pi_s)}
	&=\lim_{m\to\infty}
	\frac{
		2m\left(1+\frac{m-1}{m^2}\right)-1
	}{
		2\left(2-\frac{1}{m}\right)^2-1
	}&&\\
	&=\lim_{m\to\infty}
	\frac{2m+1-\frac{2}{m}}
	{7-\frac{8}{m}+\frac{2}{m^2}}&&\\
	&=\infty.&&
\end{flalign*}
Thus, sufficiently large but finite numbers of
servers and users can yield arbitrarily large
equilibrium cost ratios. Hence, the PoA is unbounded.
 \end{proof}

\section{Solution to the ToC: Penalty Mechanism}\label{sec:soltoc}

The preceding analysis shows that selfish routing places excessive
cumulative load on faster servers. We now consider a penalty mechanism
that changes each user's routing objective while retaining decentralized
decision making. The system, feasible strategies, and queueing assumptions
remain the same as in the distributed model.

Following the motivation for decentralized management of common
resources~\cite{ostrom_1990}, we assign a nonnegative penalty factor
$\sigma_j$ to each server $j\in[m]$. Specifically, let
$\sigma_j=\beta\Delta_j$, where
$\Delta_j=\mu_j/\min_{r\in[m]}\mu_r$ and $\beta\geq0$.
Since the servers are ordered by service rate, the penalty factors
satisfy $\sigma_1\geq\cdots\geq\sigma_m\geq0$.
A user incurs the penalty only on the fraction of its jobs routed
to the corresponding server. Thus,
\begin{equation}\label{eq:penalty-payment}
	\mathcal{P}_i(\mathbf{x}_i,\mathbf{x}_{-i})
	=\sum_{j=1}^{m}\sigma_j
	\frac{x_{ij}}{\mu_j^i-\phi_i x_{ij}}.
\end{equation}
The penalized objective of player $i$ is
\begin{equation}\label{eq:penalty-objective}
	\overline{\mathcal{R}}_i(\mathbf{x}_i,\mathbf{x}_{-i})
	=\mathcal{R}_i(\mathbf{x}_i,\mathbf{x}_{-i})
	+\mathcal{P}_i(\mathbf{x}_i,\mathbf{x}_{-i})
	=\sum_{j=1}^{m}(1+\sigma_j)
	\frac{x_{ij}}{\mu_j^i-\phi_i x_{ij}}.
\end{equation}
Here, $\sigma_j$ is a dimensionless weight on response time.
The penalty therefore changes routing incentives but does not
alter the service rates $\mu_j$.

\subsection{Penalty-Augmented Resource Allocation Game}

\begin{defn}[Penalty-augmented resource allocation game]
	\label{def:penalty-game}
	Define the game by
	\[
	\overline{\Gamma}=\langle[n],(\mathcal{S}_i)_{i\in[n]},
	(\overline{\mathcal{R}}_i)_{i\in[n]}\rangle,
	\]
	where the strategy sets $\mathcal{S}_i$ are the same as in
	Definition~\ref{def:game}, subject to the system feasibility
	constraints, and $\overline{\mathcal{R}}_i$ is defined in
	\eqref{eq:penalty-objective}.
\end{defn}

A feasible profile $\mathbf{X}^{*}=(\mathbf{x}_1^{*},\ldots,
\mathbf{x}_n^{*})$ is an NE of $\overline{\Gamma}$ if, for every
$i\in[n]$,
\[
\mathbf{x}_i^{*}\in
\argmin_{\mathbf{x}_i\in\mathcal{S}_i(\mathbf{x}_{-i}^{*})}
\overline{\mathcal{R}}_i(\mathbf{x}_i,\mathbf{x}_{-i}^{*}).
\]
When $\beta=0$, this game coincides with $\Gamma$.

\subsection{Characterizing the Penalty-Based Best Response}

\begin{thm}\label{thm:penaltycharc}
	Fix $\mathbf{x}_{-i}$ and suppose the individual feasible set
	is nonempty. Then player $i$ has a unique minimizer of
	\eqref{eq:penalty-objective}, given by
	\begin{equation}\label{eq:penalty-best-response}
		x_{ij}=\frac{1}{\phi_i}
		\left[\mu_j^i-
		\sqrt{\frac{(1+\sigma_j)\mu_j^i}{\overline{\alpha}_i}}
		\right]_+,\qquad j\in[m],
	\end{equation}
	where $\overline{\alpha}_i>0$ is uniquely chosen so that
	$\sum_{j=1}^{m}x_{ij}=1$. Server $j$ is used by player $i$
	if and only if $\overline{\alpha}_i>(1+\sigma_j)/\mu_j^i$.
\end{thm}

\begin{proof}
	For each $j\in[m]$, the first and second derivatives
	of the corresponding term in \eqref{eq:penalty-objective} are
	\begin{flalign*}
		&\frac{(1+\sigma_j)\mu_j^i}
		{(\mu_j^i-\phi_i x_{ij})^2}>0,&&\\
		&\frac{2(1+\sigma_j)\mu_j^i\phi_i}
		{(\mu_j^i-\phi_i x_{ij})^3}>0.&&
	\end{flalign*}
	The objective is therefore strictly convex and
	approaches infinity as any server load approaches
	its capacity. Hence, a unique minimizer exists.
	The KKT conditions give
	\[
	\frac{(1+\sigma_j)\mu_j^i}
	{(\mu_j^i-\phi_i x_{ij})^2}
	\begin{cases}
		=\overline{\alpha}_i,&x_{ij}>0,\\
		\geq\overline{\alpha}_i,&x_{ij}=0.
	\end{cases}
	\]

Solving these conditions yields
\eqref{eq:penalty-best-response}, where
$\overline{\alpha}_i$ is uniquely determined
by the requirement that the allocation fractions
are nonnegative and sum to one.
\end{proof}

As in Fact~\ref{fact:1}, individual optimality must hold
simultaneously at a penalized equilibrium. We express these
conditions using the actual aggregate server loads.

\begin{fact}\label{fact:3}
	Let $\mathbf{X}^{*}$ be a feasible NE of $\overline{\Gamma}$,
	with $\lambda_j^*=\sum_{i=1}^{n}\phi_i x_{ij}^{*}$.
	For every player $i\in[n]$, there exists
	$\overline{\alpha}_i>0$ such that
	\begin{equation}\label{eq:fact31}
		(1+\sigma_j)\left(
		\frac{1}{\mu_j-\lambda_j^*}
		+\frac{\phi_i x_{ij}^{*}}{(\mu_j-\lambda_j^*)^2}
		\right)
		\begin{cases}
			=\overline{\alpha}_i,&x_{ij}^{*}>0,\\
			\geq\overline{\alpha}_i,&x_{ij}^{*}=0,
		\end{cases}
		\quad j\in[m].
	\end{equation}
	Moreover,
	\begin{equation}\label{eq:fact32}
		\sum_{j=1}^{m}x_{ij}^{*}=1\quad(i\in[n]),
		\qquad\sum_{j=1}^{m}\lambda_j^*=\Phi,
		\qquad\lambda_j^*<\mu_j\quad(j\in[m]).
	\end{equation}
	Conversely, any feasible profile satisfying these
	conditions for all players is an NE of $\overline{\Gamma}$.
\end{fact}

\begin{proof}
	At an equilibrium,
	$\mu_j^i-\phi_i x_{ij}^{*}=\mu_j-\lambda_j^p$.
	Differentiating \eqref{eq:penalty-objective} and applying
	the KKT conditions yields \eqref{eq:fact31}.
	Equation~\eqref{eq:fact32} follows from feasibility.
	Conversely, strict convexity makes these conditions
	sufficient for each player's best response.
\end{proof}

\subsection{Theoretical Analysis of the Penalty Mechanism}

We now compare the original and penalized equilibria. For the
comparison of aggregate loads, assume that all players have the
same arrival rate $\phi_i=\Phi/n$, as in
Theorem~\ref{thm:symmetric-ne}. This assumption ensures that
the symmetric equilibria of both games can be represented by
one multiplier per game. Without it, the player-specific
multipliers in Facts~\ref{fact:1} and \ref{fact:3}
do not directly give a load ordering.

\begin{lemma}\label{lemma:4}
	Suppose $\phi_i=\Phi/n$ for every $i\in[n]$ and
	$0<\Phi<\sum_{j=1}^{m}\mu_j$. The penalized game
	admits a unique symmetric NE. Its active servers form
	a prefix $\overline{\mathcal{K}}=\{1,\ldots,\overline{k}\}$.
\end{lemma}

\begin{proof}
	At a symmetric profile, player $i$ contributes
	$\lambda_j^p/n$ to server $j$. Consequently,
	\eqref{eq:fact31} becomes
	\begin{flalign*}
		&(1+\sigma_j)\left(
		\frac{1}{\mu_j-\lambda_j^p}
		+\frac{\lambda_j^p}{n(\mu_j-\lambda_j^p)^2}
		\right)=\overline{\alpha}
		&&
	\end{flalign*}
	for each active server; an inactive server satisfies
	$\overline{\alpha}\leq(1+\sigma_j)/\mu_j$.
	The left-hand side is continuous and strictly increasing
	in $\lambda_j^p\in[0,\mu_j)$. Thus, for each
	$\overline{\alpha}>0$, the corresponding load is uniquely
	determined, and it increases continuously from zero
	toward $\mu_j$ as $\overline{\alpha}$ increases.
	Since their sum must equal $\Phi$, there is a unique
	$\overline{\alpha}$ and hence a unique symmetric NE.
	Finally,
	\[
	\frac{1+\sigma_j}{\mu_j}
	=\frac{1}{\mu_j}
	+\frac{\beta}{\min_{r\in[m]}\mu_r}
	\]
	is nondecreasing in $j$.
	Therefore, the active servers consist of the
	servers with the highest service rates, forming
	a prefix of the ordered server set.	
\end{proof}

\begin{lemma}\label{lemma:5}
	Under the assumptions of Lemma~\ref{lemma:4},
	$\mathcal{K}_d\subseteq\overline{\mathcal{K}}$.
	In particular, $k_d\leq\overline{k}$.
\end{lemma}

\begin{proof}
	Let $\alpha^d$ and $\overline{\alpha}$ be the common
	multipliers at the symmetric equilibria of $\Gamma$
	and $\overline{\Gamma}$, respectively. Suppose, to the
	contrary, that $\overline{k}<k_d$. Since server
	$\overline{k}+1$ is inactive under the penalty mechanism
	but active at the original NE,
	\[
	\frac{\overline{\alpha}}{1+\sigma_{\overline{k}+1}}
	\leq\frac{1}{\mu_{\overline{k}+1}}<\alpha^d.
	\]
	As $\sigma_j\geq\sigma_{\overline{k}+1}$ for
	$j\leq\overline{k}$, we also have
	$\overline{\alpha}/(1+\sigma_j)<\alpha^d$.
	The marginal response-time expression is strictly
	increasing in server load. Hence,
	$\lambda_j^p<\lambda_j^d$ for every
	$j\leq\overline{k}$. For $j>\overline{k}$,
	$\lambda_j^p=0\leq\lambda_j^d$.
	This contradicts
	$\sum_{j=1}^{m}\lambda_j^p
	=\sum_{j=1}^{m}\lambda_j^d=\Phi$.
	Therefore, $\overline{k}\geq k_d$.
\end{proof}

\begin{thm}\label{thm:final}
	Under the assumptions of Lemma~\ref{lemma:4},
	there exists an index $\ell\in\{1,\ldots,m+1\}$
	such that
	\begin{enumerate}[label=(\roman*)]
		\item $\lambda_j^p\leq\lambda_j^d$ for every $j<\ell$;
		\item $\lambda_j^p\geq\lambda_j^d$ for every $j\geq\ell$.
	\end{enumerate}
	Consequently, for every $r\in[m]$,
	\begin{equation}\label{eq:penalty-prefix-load}
		\sum_{j=1}^{r}\lambda_j^p
		\leq\sum_{j=1}^{r}\lambda_j^d.
	\end{equation}
\end{thm}

\begin{proof}
	Let $\alpha^d$ and $\overline{\alpha}$ denote the
	common multipliers at the two symmetric equilibria.
	By the marginal conditions in Fact~\ref{fact:1}
	and Lemma~\ref{lemma:4}, the assigned load on
	server $j$ increases with the effective multiplier.
	The effective multiplier is $\alpha^d$ at the
	original NE and $\overline{\alpha}/(1+\sigma_j)$
	at the penalized NE.
	
	\noindent\textbf{Part (i):}
	Since $\sigma_1\geq\cdots\geq\sigma_m$,
	$\overline{\alpha}/(1+\sigma_j)$ is nondecreasing
	in $j$. Let $\ell$ be the smallest index for which
	$\overline{\alpha}/(1+\sigma_\ell)\geq\alpha^d$.
	If no such index exists, let $\ell=m+1$.
	For every $j<\ell$, we have
	$\overline{\alpha}/(1+\sigma_j)<\alpha^d$,
	which implies $\lambda_j^p\leq\lambda_j^d$.
	
	\noindent\textbf{Part (ii):}
	For every $j\geq\ell$,
	$\overline{\alpha}/(1+\sigma_j)\geq\alpha^d$.
	The same marginal conditions give
	$\lambda_j^p\geq\lambda_j^d$.
	
	For $r<\ell$, Part (i) directly implies
	\eqref{eq:penalty-prefix-load}. For $r\geq\ell$,
	Part (ii) gives
	$\sum_{j=r+1}^{m}\lambda_j^p
	\geq\sum_{j=r+1}^{m}\lambda_j^d$.
	Since both allocations have total load $\Phi$,
	\begin{flalign*}
		\sum_{j=1}^{r}\lambda_j^p
		=&~\Phi-\sum_{j=r+1}^{m}\lambda_j^p\\
		\leq&~\Phi-\sum_{j=r+1}^{m}\lambda_j^d\\
		=&~\sum_{j=1}^{r}\lambda_j^d.
	\end{flalign*}
	Thus, \eqref{eq:penalty-prefix-load} holds for
	every $r\in[m]$.
\end{proof}

\begin{remark}
	Theorem~\ref{thm:final} establishes that the penalty
	mechanism shifts cumulative load away from the faster
	servers in the symmetric-user setting. It does not
	establish that the penalized allocation coincides with
	the social optimum, nor does it by itself imply a
	reduction in total system cost.
\end{remark}

\section{Effect of Penalty Mechanism on Inefficiency}
\label{sec:penaltypoa}

We now examine whether the change in routing decisions
also reduces the inefficiency measured in Section~\ref{sec:quantpoa}.
Let $\pi_d$, $\pi_p$, and $\pi_s$ denote the original
NE, the penalized NE, and the socially optimal allocation,
respectively. Their aggregate server loads are
$\lambda_j^d$, $\lambda_j^p$, and $\lambda_j^s$.
The penalty in \eqref{eq:penalty-payment} is a routing
incentive, not an additional server processing delay.
Accordingly, we evaluate the physical system cost using
\eqref{eq:poacostmdl}, without adding penalty transfers
to the congestion cost.

For the selected equilibria, let
\[
\PoA_d=\frac{C(\pi_d)}{C(\pi_s)},\qquad
\PoA_p=\frac{C(\pi_p)}{C(\pi_s)}.
\]
These quantities are equilibrium cost ratios. The
worst-case PoA of each game requires taking the supremum
over its Nash equilibria, as in the PoA definition.

Theorem~\ref{thm:final} gives a structural load comparison
for symmetric users. A reduction in total system cost,
however, requires a further condition on how far the
penalized allocation moves toward the social optimum.

\begin{lemma}\label{lem:pen-cost-reduction}
	If there exists $\theta\in[0,1]$ such that
	$
	\lambda_j^p=(1-\theta)\lambda_j^d
	+\theta\lambda_j^s, j\in[m].
	$
	Then
	\begin{equation}\label{eq:penalty-cost-convex}
		C(\pi_p)\leq(1-\theta)C(\pi_d)
		+\theta C(\pi_s).
	\end{equation}
	If $0<\theta<1$ and
	$(\lambda_1^d,\ldots,\lambda_m^d)
	\neq(\lambda_1^s,\ldots,\lambda_m^s)$,
	the inequality is strict.
\end{lemma}

\begin{proof}
	For $0\leq\lambda_j<\mu_j$, the function
	$\lambda_j/(\mu_j-\lambda_j)$ is strictly convex,
	since
	\[
	\frac{\mathrm{d}^2}{\mathrm{d}\lambda_j^2}
	\left(\frac{\lambda_j}{\mu_j-\lambda_j}\right)
	=\frac{2\mu_j}{(\mu_j-\lambda_j)^3}>0.
	\]
	Applying convexity to the stated combination of
	server loads and summing over $j\in[m]$ yields
	\eqref{eq:penalty-cost-convex}. Strict convexity
	gives the strict inequality under the stated conditions.
\end{proof}

\begin{thm}\label{thm:penalty-poa}
	Under the conditions of Lemma~\ref{lem:pen-cost-reduction},
	\begin{equation}\label{eq:penalty-poa-bound}
		1\leq\PoA_p\leq(1-\theta)\PoA_d+\theta
		\leq\PoA_d,
		\qquad
		\PoA_d-\PoA_p\geq\theta(\PoA_d-1).
	\end{equation}
	If $\theta>0$ and $\PoA_d>1$, then
	$\PoA_p<\PoA_d$.
\end{thm}

\begin{proof}
	Since $\pi_s$ is the socially optimal allocation,
	its physical system cost cannot exceed that of
	the penalized equilibrium $\pi_p$. Thus,
	$C(\pi_s)\leq C(\pi_p)$, which implies
	$\PoA_p\geq1$. Furthermore, dividing
	\eqref{eq:penalty-cost-convex} by $C(\pi_s)>0$
	yields
$
	1\leq\PoA_p\leq(1-\theta)\PoA_d+\theta.
$
	Since $\PoA_d\geq1$ and $0\leq\theta\leq1$,
	we have
	$(1-\theta)\PoA_d+\theta\leq\PoA_d$.
	Rearranging the upper bound also gives
$
	\PoA_d-\PoA_p\geq\theta(\PoA_d-1).
$
	Finally, if $\theta>0$ and $\PoA_d>1$,
	then $\theta(\PoA_d-1)>0$, implying
	$\PoA_p<\PoA_d$. This completes the proof.
\end{proof}

\begin{remark}
	The improvement in Theorem~\ref{thm:penalty-poa}
	is conditional. A positive penalty does not by itself
	imply that the penalized equilibrium loads lie on the
	line segment joining the original equilibrium and the
	social optimum. Thus, Theorem~\ref{thm:final} cannot
	be used alone to conclude that the system cost or PoA
	always decreases as $\beta$ increases.
\end{remark}

\subsection{Discussion on the Proposed Penalty Mechanism}

The penalty mechanism modifies the individual routing
objective by assigning larger delay weights to faster
servers. Theorem~\ref{thm:penaltycharc} characterizes
each player's unique best response, while
Fact~\ref{fact:3} gives the coupled equilibrium conditions.
For identical user arrival rates,
Lemmas~\ref{lemma:4} and \ref{lemma:5} and
Theorem~\ref{thm:final} show that the penalized
equilibrium uses at least as many servers as the
unpenalized equilibrium and moves cumulative load
toward slower servers.

This redistribution addresses the load-concentration
aspect of the ToC identified in
Theorem~\ref{thm:main_inefficiency}. However, a shift
toward slower servers does not necessarily reduce
total congestion: an excessively large penalty can
move more load than is socially desirable. Under the
load-interpolation condition of
Lemma~\ref{lem:pen-cost-reduction}, the social cost
decreases and Theorem~\ref{thm:penalty-poa} gives
a corresponding improvement in the equilibrium cost ratio.

A limitation of the proposed penalty mechanism is
that its effectiveness depends on the choice of
$\beta$. Furthermore, the theoretical guarantee
of improved system efficiency is not established
for arbitrary heterogeneous players. Extending
the analysis to such settings and determining
suitable penalty parameters constitute promising
directions for future research.

\section{Conclusion and Future Work}\label{sec:con}

This paper investigated the Tragedy of the Commons (ToC) in heterogeneous
distributed resource allocation from an algorithmic game-theoretic
perspective. To this end, we formulated the distributed resource allocation problem as a non-cooperative game and
derived closed-form analytical characterizations of both the
distributed Nash equilibrium and the socially optimal allocation.
Building upon these results, we established the structural conditions
under which selfish users overload high-capacity servers, thereby
establishing the emergence of the ToC in heterogeneous distributed systems.

Using these structural characterizations, we derived an exact
analytical expression for the PoA and investigated its asymptotic behavior under varying traffic conditions. The analysis showed that the inefficiency introduced by selfish routing can become arbitrarily large under suitable system configurations. These results characterized equilibrium inefficiency and revealed the relationship between server heterogeneity, equilibrium structure, and overall system performance.

Motivated by these observations, we proposed a penalty-based incentive
mechanism for mitigating the ToC while preserving the decentralized nature of users' decisions. 
For symmetric users, we established how the
penalty mechanism redistributes equilibrium
loads. We also derived sufficient conditions
under which it reduces the overall system cost
and improves equilibrium efficiency. Overall, the analytical framework developed in this paper unifies equilibrium analysis, the PoA, and incentive-based mechanism design within a common mathematical framework for heterogeneous distributed resource allocation.

Several directions remain for future research. An immediate extension
is to generalize the proposed framework to stochastic settings with
uncertain arrivals, time-varying server capacities, and incomplete
information. Another interesting direction is to develop adaptive
penalty mechanisms that dynamically adjust incentives according to the
system state. It would also be worthwhile to extend the analysis to
more general resource allocation problems, such as network routing,
cloud-edge systems, federated learning, and multi-resource allocation.
Finally, extending the proposed framework to dynamic games and online
resource allocation remains an interesting direction for future work.

\section*{CRediT authorship contribution statement}

\textbf{Avadh Kishor:} Conceptualization,
Methodology, Formal analysis, Writing -- original draft. \textbf{Rajdeep Niyogi:} Supervision,
Validation, Writing -- review \& editing.


\bibliographystyle{model1-num-names}
%\bibliographystyle{unsrt}

\bibliography{toc-tpds}

\begin{thebibliography}{36}
\expandafter\ifx\csname natexlab\endcsname\relax\def\natexlab#1{#1}\fi
\providecommand{\url}[1]{\texttt{#1}}
\providecommand{\href}[2]{#2}
\providecommand{\path}[1]{#1}
\providecommand{\DOIprefix}{doi:}
\providecommand{\ArXivprefix}{arXiv:}
\providecommand{\URLprefix}{URL: }
\providecommand{\Pubmedprefix}{pmid:}
\providecommand{\doi}[1]{\href{http://dx.doi.org/#1}{\path{#1}}}
\providecommand{\Pubmed}[1]{\href{pmid:#1}{\path{#1}}}
\providecommand{\bibinfo}[2]{#2}
\ifx\xfnm\relax \def\xfnm[#1]{\unskip,\space#1}\fi
%Type = Book
\bibitem[{Nisan et~al.(2007)Nisan, Roughgarden, Tardos, and
  Vazirani}]{nisan2007algorithmic}
\bibinfo{editor}{N.~Nisan}, \bibinfo{editor}{T.~Roughgarden},
  \bibinfo{editor}{E.~Tardos}, \bibinfo{editor}{V.~V. Vazirani} (Eds.),
  \bibinfo{title}{Algorithmic Game Theory}, \bibinfo{publisher}{Cambridge
  University Press}, \bibinfo{year}{2007}.
%Type = Article
\bibitem[{Harks et~al.(2022)}]{harks2022equilibrium}
\bibinfo{author}{T.~Harks}, et~al.,
\newblock \bibinfo{title}{Equilibrium computation in resource allocation
  games},
\newblock \bibinfo{journal}{Mathematical Programming}  (\bibinfo{year}{2022}).
%Type = Article
\bibitem[{Shamshirband et~al.(2021)Shamshirband, Joloudari, Shirkharkolaie,
  Mojrian, Rahmani, Mostafavi, and Mansor}]{shamshirband2021survey}
\bibinfo{author}{S.~Shamshirband}, \bibinfo{author}{J.~H. Joloudari},
  \bibinfo{author}{S.~K. Shirkharkolaie}, \bibinfo{author}{S.~Mojrian},
  \bibinfo{author}{F.~Rahmani}, \bibinfo{author}{S.~Mostafavi},
  \bibinfo{author}{Z.~Mansor},
\newblock \bibinfo{title}{Game theory and evolutionary optimization approaches
  applied to resource allocation problems in computing environments: A survey},
\newblock \bibinfo{journal}{Mathematical Biosciences and Engineering}
  \bibinfo{volume}{18} (\bibinfo{year}{2021}) \bibinfo{pages}{9190--9232}.
%Type = Article
\bibitem[{Grosu and Chronopoulos(2005)}]{grosu2005noncooperative}
\bibinfo{author}{D.~Grosu}, \bibinfo{author}{A.~T. Chronopoulos},
\newblock \bibinfo{title}{Noncooperative load balancing in distributed
  systems},
\newblock \bibinfo{journal}{Journal of parallel and distributed computing}
  \bibinfo{volume}{65} (\bibinfo{year}{2005}) \bibinfo{pages}{1022--1034}.
%Type = Article
\bibitem[{Penmatsa and Chronopoulos(2011)}]{penmatsa2011static}
\bibinfo{author}{S.~Penmatsa}, \bibinfo{author}{A.~T. Chronopoulos},
\newblock \bibinfo{title}{Game-theoretic static load balancing for distributed
  systems},
\newblock \bibinfo{journal}{Journal of Parallel and Distributed Computing}
  \bibinfo{volume}{71} (\bibinfo{year}{2011}) \bibinfo{pages}{537--555}.
%Type = Article
\bibitem[{L{\'o}pez et~al.(2005)L{\'o}pez, del Rey~Almansa, Paquelet, and
  Fern{\'a}ndez}]{lopez2005mathematical}
\bibinfo{author}{L.~L{\'o}pez}, \bibinfo{author}{G.~del Rey~Almansa},
  \bibinfo{author}{S.~Paquelet}, \bibinfo{author}{A.~Fern{\'a}ndez},
\newblock \bibinfo{title}{A mathematical model for the tcp tragedy of the
  commons},
\newblock \bibinfo{journal}{Theoretical Computer Science} \bibinfo{volume}{343}
  (\bibinfo{year}{2005}) \bibinfo{pages}{4--26}.
%Type = Inproceedings
\bibitem[{Saha and Sen(2003)}]{saha2003local}
\bibinfo{author}{S.~Saha}, \bibinfo{author}{S.~Sen},
\newblock \bibinfo{title}{Local decision procedures for avoiding the tragedy of
  commons},
\newblock in: \bibinfo{booktitle}{International Workshop on Distributed
  Computing}, \bibinfo{organization}{Springer}, \bibinfo{year}{2003}, pp.
  \bibinfo{pages}{311--320}.
%Type = Inproceedings
\bibitem[{Kishor et~al.(2016)Kishor, Gargt, and Niyogi}]{kishor2016altruistic}
\bibinfo{author}{A.~Kishor}, \bibinfo{author}{T.~Gargt},
  \bibinfo{author}{R.~Niyogi},
\newblock \bibinfo{title}{Altruistic decision making approach to resolve the
  tragedy of the commons},
\newblock in: \bibinfo{booktitle}{2016 International Conference on Advances in
  Computing, Communications and Informatics (ICACCI)},
  \bibinfo{organization}{IEEE}, \bibinfo{year}{2016}, pp.
  \bibinfo{pages}{1858--1863}.
%Type = Article
\bibitem[{Roughgarden and Tardos(2002)}]{roughgarden2002bad}
\bibinfo{author}{T.~Roughgarden}, \bibinfo{author}{{\'E}.~Tardos},
\newblock \bibinfo{title}{How bad is selfish routing?},
\newblock \bibinfo{journal}{Journal of the ACM (JACM)} \bibinfo{volume}{49}
  (\bibinfo{year}{2002}) \bibinfo{pages}{236--259}.
%Type = Inproceedings
\bibitem[{Christodoulou and Koutsoupias(2005)}]{christodoulou2005price}
\bibinfo{author}{G.~Christodoulou}, \bibinfo{author}{E.~Koutsoupias},
\newblock \bibinfo{title}{The price of anarchy of finite congestion games},
\newblock in: \bibinfo{booktitle}{Proceedings of the thirty-seventh annual ACM
  symposium on Theory of computing}, \bibinfo{year}{2005}, pp.
  \bibinfo{pages}{67--73}.
%Type = Article
\bibitem[{Chen et~al.(2020)Chen, Hu, Wang, and Wu}]{chen2020randomscheduler}
\bibinfo{author}{X.~Chen}, \bibinfo{author}{X.~Hu}, \bibinfo{author}{C.~Wang},
  \bibinfo{author}{X.~Wu},
\newblock \bibinfo{title}{The efficiency of nash equilibria in the load
  balancing game with a randomizing scheduler},
\newblock \bibinfo{journal}{Theoretical Computer Science} \bibinfo{volume}{838}
  (\bibinfo{year}{2020}) \bibinfo{pages}{180--194}.
%Type = Article
\bibitem[{Paccagnan et~al.(2020)Paccagnan, Chandan, and
  Marden}]{paccagnan2019utility}
\bibinfo{author}{D.~Paccagnan}, \bibinfo{author}{R.~Chandan},
  \bibinfo{author}{J.~R. Marden},
\newblock \bibinfo{title}{Utility design for distributed resource allocation --
  part i: Characterizing and optimizing the exact price of anarchy},
\newblock \bibinfo{journal}{IEEE Transactions on Automatic Control}
  \bibinfo{volume}{65} (\bibinfo{year}{2020}) \bibinfo{pages}{4616--4631}.
%Type = Article
\bibitem[{Jiang(2015)}]{jiang2015survey}
\bibinfo{author}{Y.~Jiang},
\newblock \bibinfo{title}{A survey of task allocation and load balancing in
  distributed systems},
\newblock \bibinfo{journal}{IEEE Transactions on Parallel and Distributed
  Systems} \bibinfo{volume}{27} (\bibinfo{year}{2015})
  \bibinfo{pages}{585--599}.
%Type = Article
\bibitem[{Gamez et~al.(2018)Gamez, Camacho, and
  Rodríguez}]{gamez2018marketing}
\bibinfo{author}{M.~Gamez}, \bibinfo{author}{J.~F. Camacho},
  \bibinfo{author}{J.~A. Rodríguez},
\newblock \bibinfo{title}{An evolutionary game-theoretic approach to marketing
  cooperatives in fisheries},
\newblock \bibinfo{journal}{Applied Mathematics and Computation}
  \bibinfo{volume}{329} (\bibinfo{year}{2018}) \bibinfo{pages}{165--176}.
%Type = Article
\bibitem[{Grosu and Chronopoulos(2004)}]{grosu2004mechanism}
\bibinfo{author}{D.~Grosu}, \bibinfo{author}{A.~T. Chronopoulos},
\newblock \bibinfo{title}{A truthful mechanism for fair load balancing in
  distributed systems},
\newblock \bibinfo{journal}{Parallel and Distributed Processing Symposium
  (IPDPS)}  (\bibinfo{year}{2004}) \bibinfo{pages}{1--8}.
%Type = Article
\bibitem[{Polgar and Varga(2023)}]{polgar2023loadbalancing}
\bibinfo{author}{Z.~A. Polgar}, \bibinfo{author}{M.~Varga},
\newblock \bibinfo{title}{Game theory-based load-balancing algorithms for small
  cells wireless backhaul connections},
\newblock \bibinfo{journal}{Applied Sciences} \bibinfo{volume}{13}
  (\bibinfo{year}{2023}) \bibinfo{pages}{1485}.
%Type = Article
\bibitem[{Kishor et~al.(2020{\natexlab{a}})Kishor, Niyogi, and
  Veeravalli}]{avadh20}
\bibinfo{author}{A.~Kishor}, \bibinfo{author}{R.~Niyogi},
  \bibinfo{author}{B.~Veeravalli},
\newblock \bibinfo{title}{Fairness-aware mechanism for load balancing in
  distributed systems},
\newblock \bibinfo{journal}{IEEE Transactions on Services Computing}
  \bibinfo{volume}{Early access} (\bibinfo{year}{2020}{\natexlab{a}})
  \bibinfo{pages}{1--14}.
%Type = Article
\bibitem[{Kishor et~al.(2020{\natexlab{b}})Kishor, Niyogi, and
  Veeravalli}]{KISHOR202029}
\bibinfo{author}{A.~Kishor}, \bibinfo{author}{R.~Niyogi},
  \bibinfo{author}{B.~Veeravalli},
\newblock \bibinfo{title}{A game-theoretic approach for cost-aware load
  balancing in distributed systems},
\newblock \bibinfo{journal}{Future Generation Computer Systems}
  \bibinfo{volume}{109} (\bibinfo{year}{2020}{\natexlab{b}}) \bibinfo{pages}{29
  -- 44}.
%Type = Book
\bibitem[{Turner(1993)}]{turner1993tragedy}
\bibinfo{author}{R.~M. Turner}, \bibinfo{title}{The tragedy of the commons and
  distributed AI systems}, \bibinfo{publisher}{Department of Computer Science,
  University of New Hampshire Durham, NH}, \bibinfo{year}{1993}.
%Type = Inproceedings
\bibitem[{Mundhe and Sen(2000)}]{mundhe2000evolving}
\bibinfo{author}{M.~Mundhe}, \bibinfo{author}{S.~Sen},
\newblock \bibinfo{title}{Evolving agent socienties that avoid social
  dilemmas.},
\newblock in: \bibinfo{booktitle}{GECCO}, \bibinfo{year}{2000}, pp.
  \bibinfo{pages}{809--816}.
%Type = Inproceedings
\bibitem[{Tumer and Wolpert(2000)}]{tumer2000collective}
\bibinfo{author}{K.~Tumer}, \bibinfo{author}{D.~H. Wolpert},
\newblock \bibinfo{title}{Collective intelligence and braess' paradox},
\newblock in: \bibinfo{booktitle}{Aaai/iaai}, \bibinfo{year}{2000}, pp.
  \bibinfo{pages}{104--109}.
%Type = Article
\bibitem[{Płatkowski(2016)}]{platkowski2016egalitarian}
\bibinfo{author}{T.~Płatkowski},
\newblock \bibinfo{title}{Egalitarian redistribution in multiplayer social
  dilemmas},
\newblock \bibinfo{journal}{Applied Mathematics and Computation}
  \bibinfo{volume}{282} (\bibinfo{year}{2016}) \bibinfo{pages}{139--151}.
%Type = Article
\bibitem[{Płatkowski(2017)}]{platkowski2017greed}
\bibinfo{author}{T.~Płatkowski},
\newblock \bibinfo{title}{Greed and fear in evolutionary multiplayer social
  dilemmas},
\newblock \bibinfo{journal}{Applied Mathematics and Computation}
  \bibinfo{volume}{297} (\bibinfo{year}{2017}) \bibinfo{pages}{191--201}.
%Type = Article
\bibitem[{He et~al.(2019)He, Szolnoki, and Perc}]{he2019monitoring}
\bibinfo{author}{N.~He}, \bibinfo{author}{A.~Szolnoki},
  \bibinfo{author}{M.~Perc},
\newblock \bibinfo{title}{Monitoring and reporting promote cooperation in
  collective-risk social dilemmas},
\newblock \bibinfo{journal}{Applied Mathematics and Computation}
  \bibinfo{volume}{347} (\bibinfo{year}{2019}) \bibinfo{pages}{773--782}.
%Type = Article
\bibitem[{Gao et~al.(2019)Gao, Szolnoki, and Perc}]{gao2019redistribution}
\bibinfo{author}{J.~Gao}, \bibinfo{author}{A.~Szolnoki},
  \bibinfo{author}{M.~Perc},
\newblock \bibinfo{title}{Asymmetric payoff redistribution promotes cooperation
  in structured populations},
\newblock \bibinfo{journal}{Applied Mathematics and Computation}
  \bibinfo{volume}{346} (\bibinfo{year}{2019}) \bibinfo{pages}{295--305}.
%Type = Article
\bibitem[{Jiang et~al.(2021)Jiang, Szolnoki, and Perc}]{jiang2021bystander}
\bibinfo{author}{L.~Jiang}, \bibinfo{author}{A.~Szolnoki},
  \bibinfo{author}{M.~Perc},
\newblock \bibinfo{title}{The bystander effect in collective-risk social
  dilemmas},
\newblock \bibinfo{journal}{Applied Mathematics and Computation}
  \bibinfo{volume}{402} (\bibinfo{year}{2021}) \bibinfo{pages}{126152}.
%Type = Article
\bibitem[{Su et~al.(2023)Su, Szolnoki, and Perc}]{su2023interdependent}
\bibinfo{author}{X.~Su}, \bibinfo{author}{A.~Szolnoki},
  \bibinfo{author}{M.~Perc},
\newblock \bibinfo{title}{Cooperation on interdependent networks with resource
  interaction},
\newblock \bibinfo{journal}{Applied Mathematics and Computation}
  \bibinfo{volume}{446} (\bibinfo{year}{2023}) \bibinfo{pages}{127887}.
%Type = Article
\bibitem[{Mehta and Vazirani(2020)}]{mehta2020market}
\bibinfo{author}{R.~Mehta}, \bibinfo{author}{V.~V. Vazirani},
\newblock \bibinfo{title}{An incentive compatible, efficient market for air
  traffic flow management},
\newblock \bibinfo{journal}{Theoretical Computer Science} \bibinfo{volume}{818}
  (\bibinfo{year}{2020}) \bibinfo{pages}{41--50}.
%Type = Article
\bibitem[{Chen et~al.(2021)Chen, Gu, and Wang}]{chen2021fair}
\bibinfo{author}{Z.~Chen}, \bibinfo{author}{Z.~Gu}, \bibinfo{author}{Y.~Wang},
\newblock \bibinfo{title}{Which is the fairest allocation in the max-min
  fairness-based coalitional game?},
\newblock \bibinfo{journal}{Theoretical Computer Science} \bibinfo{volume}{888}
  (\bibinfo{year}{2021}) \bibinfo{pages}{56--69}.
%Type = Article
\bibitem[{Gao et~al.(2025)Gao, Pan, and He}]{gao2025ecological}
\bibinfo{author}{L.~Gao}, \bibinfo{author}{Q.~Pan}, \bibinfo{author}{M.~He},
\newblock \bibinfo{title}{The effect of ecological compensation on a
  coevolutionary common pool resource game},
\newblock \bibinfo{journal}{Applied Mathematics and Computation}
  \bibinfo{volume}{490} (\bibinfo{year}{2025}) \bibinfo{pages}{129208}.
%Type = Article
\bibitem[{Lee et~al.(2023)Lee, Cleveland, and Szolnoki}]{lee2023myopic}
\bibinfo{author}{H.-W. Lee}, \bibinfo{author}{C.~Cleveland},
  \bibinfo{author}{A.~Szolnoki},
\newblock \bibinfo{title}{Restoring spatial cooperation with myopic agents in a
  three-strategy social dilemma},
\newblock \bibinfo{journal}{Applied Mathematics and Computation}
  \bibinfo{volume}{458} (\bibinfo{year}{2023}) \bibinfo{pages}{128226}.
%Type = Article
\bibitem[{Christodoulou et~al.(2014)Christodoulou, Mehlhorn, and
  Pyrga}]{christodoulou2014improving}
\bibinfo{author}{G.~Christodoulou}, \bibinfo{author}{K.~Mehlhorn},
  \bibinfo{author}{E.~Pyrga},
\newblock \bibinfo{title}{Improving the price of anarchy for selfish routing
  via coordination mechanisms},
\newblock \bibinfo{journal}{Algorithmica} \bibinfo{volume}{69}
  (\bibinfo{year}{2014}) \bibinfo{pages}{619--640}.
%Type = Article
\bibitem[{Brun and Prabhu(2016)}]{brun2016worst}
\bibinfo{author}{O.~Brun}, \bibinfo{author}{B.~Prabhu},
\newblock \bibinfo{title}{Worst-case analysis of non-cooperative load
  balancing},
\newblock \bibinfo{journal}{Annals of Operations Research}
  \bibinfo{volume}{239} (\bibinfo{year}{2016}) \bibinfo{pages}{471--495}.
%Type = Book
\bibitem[{Trivedi(1982)}]{DBLP:books/daglib/0079988}
\bibinfo{author}{K.~S. Trivedi}, \bibinfo{title}{Probability and statistics
  with reliability, queuing, and computer science applications},
  \bibinfo{publisher}{Prentice Hall}, \bibinfo{year}{1982}.
%Type = Book
\bibitem[{Boyd and Vandenberghe(2004)}]{Boyd:2004:CO:993483}
\bibinfo{author}{S.~Boyd}, \bibinfo{author}{L.~Vandenberghe},
  \bibinfo{title}{Convex Optimization}, \bibinfo{publisher}{Cambridge
  University Press}, \bibinfo{address}{New York, NY, USA},
  \bibinfo{year}{2004}.
%Type = Book
\bibitem[{Ostrom(1990)}]{ostrom_1990}
\bibinfo{author}{E.~Ostrom}, \bibinfo{title}{Governing the Commons: The
  Evolution of Institutions for Collective Action}, Political Economy of
  Institutions and Decisions, \bibinfo{publisher}{Cambridge University Press},
  \bibinfo{year}{1990}. \DOIprefix\doi{10.1017/CBO9780511807763}.

\end{thebibliography}

\end{document}